\documentclass[openany]{nature}
\usepackage[left]{lineno}
\usepackage{graphicx}
\usepackage{float}
\usepackage{amsmath}
\usepackage{amssymb}
\usepackage{subfigure}
\usepackage{cases}
\usepackage{mathrsfs}
\usepackage[flushleft]{threeparttable}
\usepackage{amssymb}
\usepackage{pifont}
\usepackage{epstopdf}
\usepackage{xcolor}
\usepackage{hyperref}
\usepackage[all]{hypcap}
\usepackage{mwe,tikz}
\usepackage{bm}
\usepackage{multirow}
\usepackage{longtable}
\usepackage{xspace}
\usepackage{rotating}
\usepackage{CJK}
\usepackage{url}
\usepackage{upgreek}
\usepackage{booktabs}
\usepackage{textcomp}
\usepackage{makecell}
\usepackage{enumitem}
\usepackage{csquotes}
\usepackage{soul}
\usepackage{tabularx}
\usepackage{overpic}
\usepackage{caption}
\usepackage{float}
\usepackage{threeparttable}
\usepackage{threeparttablex}
\usepackage{subfiles}
\usepackage{array}

\makeatletter
 
\def\emph#1 {\textit{ #1 } }

\let\saved@includegraphics\includegraphics
\AtBeginDocument{\let\includegraphics\saved@includegraphics}
\renewenvironment*{figure}{\@float{figure}}{\end@float}

\makeatother

\newcommand{\apj}{Astrophys. J.}

\newcommand{\pasp}{Publ. Astron. Soc. Pac.}
\newcommand{\apjs}{Astrophys. J. Supp.}
\newcommand{\araa}{Annu. Rev. Astron. Astrophys.}
\newcommand{\mnras}{Mon. Not. R. Astron. Soc.}
\newcommand{\apjl}{Astrophys. J. Let.}
\newcommand{\aap}{Astron. Astrophys.}
\newcommand{\aj}{Astron. J.}
\newcommand{\nat}{Nature}

\newcommand{\nar}{New Astronomy Reviews}

\definecolor{dkblue}{RGB}{54, 86, 169}

\title{A fast X-ray transient from a weak relativistic jet associated with a type Ic-BL supernova}

\author{H. Sun$^{1}$\thanks{These authors contributed equally to this work}, W.-X. Li$^{1*}$, L.-D. Liu$^{2,3*}$, H. Gao$^{4,5}$\thanks{E-mail: gaohe@bnu.edu.cn},
X.-F. Wang$^{6}$\thanks{E-mail: wang\_xf@mail.tsinghua.edu.cn},
W. Yuan$^{1,7}$\thanks{E-mail: wmy@nao.cas.cn},
B. Zhang$^{8,9}$\thanks{E-mail: bing.zhang@unlv.edu}, A. V. Filippenko$^{10}$, D. Xu$^{1,11}$, T. An$^{12,7}$, S. Ai$^{13}$
% ...
}

\begin{document}
\captionsetup[table]{name={\bf Table}}
\captionsetup[figure]{name={\bf Fig.}}

\maketitle

\begin{affiliations}
\item{National Astronomical Observatories, Chinese Academy of Sciences, Beijing 100101, China.}
\item{Institute of Astrophysics, Central China Normal University, Wuhan 430079, China.}
\item{Key Laboratory of Quark and Lepton Physics (Central China Normal University), Ministry of Education, Wuhan 430079, China}
\item{Institute for Frontier in Astronomy and Astrophysics, Beijing Normal University, Beijing 102206, China.}
\item{School of Physics and Astronomy, Beijing Normal University, Beijing 100875, China.}
\item{Physics Department, Tsinghua University, Beijing, 100084, China.}
\item{School of Astronomy and Space Science, University of Chinese Academy of Sciences, Chinese Academy of Sciences, Beijing 100049, China.}
\item{Nevada Center for Astrophysics, University of Nevada Las Vegas, NV 89154, USA.}
\item{Department of Physics and Astronomy, University of Nevada Las Vegas, NV 89154, USA.}
\item{Department of Astronomy, University of California, Berkeley, CA 94720-3411, USA.}
\item{Altay Astronomical Observatory, Altay, Xinjiang 836500, China.}
\item{Shanghai Astronomical Observatory, Chinese Academy of Sciences (CAS), 80 Nandan Road, Shanghai 200030, China.}
\item{Niels Bohr International Academy and DARK, Niels Bohr Institute, University of Copenhagen, Blegdamsvej 17, 2100, Copenhagen, Denmark.}
\item {Department of Astronomy, Tsinghua University, Beĳing 100084, China.}
\item{Purple Mountain Observatory, Chinese Academy of Sciences, Nanjing 210023, China.}
\item{School of Astronomy and Space Science, Nanjing University, Nanjing 210093, China.}
\item{Key Laboratory of Modern Astronomy and Astrophysics (Nanjing University), Ministry of Education, China.}
\item {Cahill Center for Astrophysics, California Institute of Technology, Pasadena, CA 91125, USA.}
\item{Department of Astronomy, School of Physical Sciences, University of Science and Technology of China, Hefei 230026, China.}
\item{Department of Astronomy and Astrophysics, The Pennsylvania State University, 525 Davey Lab, University Park, PA 16802, USA.}
\item{INAF-Osservatorio Astronomico di Padova, Vicolo dell’Osservatorio 5, 35122 Padova, Italy.}
\item{Institute of Space Sciences (ICE), Consejo Superior de Investigaciones Científicas (CSIC), Barcelona, Spain.}
\item{Las Cumbres Observatory, 6740 Cortona Drive, Suite 102, Goleta, CA 93117-5575, USA.}
\item{Department of Physics, University of California, Santa Barbara, CA 93106-9530, USA.}
\item{Department of Astronomy, School of Physics, Huazhong University of Science and Technology, Wuhan  430074, China.}
\item{Ioffe Institute, Politekhnicheskaya 26, 194021 St. Petersburg, Russia.}
\item{Key Laboratory of Particle Astrophysics, Institute of High Energy Physics, Chinese Academy of Sciences, Beijing 100049, China.}
\item{Yunnan observatories, Chinese Academy of Sciences, 650011 Kunming, China.}
\item{Ulugh Beg Astronomical Institute, Uzbekistan Academy of Sciences, Tashkent, Uzbekistan.}
\item{National University of Uzbekistan, Tashkent, Uzbekistan.}
\item{Cosmic DAWN Center, Copenhagen, Denmark.} 
\item{Niels Bohr Institute, University of Copenhagen, Jagtvej 155, 2200 Copenhagen N, Denmark.}
\item{Instituto de F{\'i}sica y Astronom{\'i}a, Universidad de Valpara{\'i}so, ave. Gran Breta{\~n}a, 1111, Casilla 5030, Valpara{\'i}so, Chile.}
\item{Millennium Institute of Astrophysics, Nuncio Monse{\~n}or Sotero Sanz 100, Of. 104, Providencia, Santiago, Chile.}
\item{Centre for Astrophysics Research, University of Hertfordshire, Hatfield AL10 9AB, UK.}
\item{School of Astronomy and Space Sciences, University of Science and Technology of China, Hefei 230026, China.}
\item{National Astronomical Research Institute of Thailand, 260 Moo 4, Donkaew, Maerim, Chiang Mai, 50180, Thailand.}
\item{Chinese Academy of Sciences South America Center for Astronomy (CASSACA), National Astronomical Observatories, CAS, Beijing 100101, China.}
\item{Beijing Planetarium, Beijing Academy of Sciences and Technology, Beijing, 100044, China.}
\item{Innovation Academy for Microsatellites, Chinese Academy of Sciences, Shanghai 201210, China.}
\item{Instituto de Astrof\'isica de Andaluc\'ia (IAA-CSIC), Glorieta de la Astronom\'ia s/n, 18008 Granada, Spain.}
\item{Ingeniería de Sistemas y Autom\'atica, Universidad de M\'alaga, Unidad Asociada al CSIC por el IAA, Escuela de Ingenier\'ias Industriales, Arquitecto Francisco Pe\~nalosa, 6, Campanillas, 29071 M\'alaga, Spain.}
\item{Shanghai Institute of Technical Physics, Chinese Academy of Sciences, Shanghai 200083, China.}
\item{Key Laboratory of Technology on Space Energy Conversion, Technical Institute of Physics and Chemistry, CAS, Beijing 100190, China.}
\item{CEA Paris-Saclay, IRFU/Département d’Astrophysique-AIM, 91191 Gif-sur-Yvette, France.}
\item{North Night Vision Technology Co., LTD, Nanjing, China.}
\item{European Space Agency, ESTEC, Keplerlaan 1, NL-2200 AG, Noordwijk, The Netherlands.}
\item{Max-Planck-Institut für extraterrestrische Physik, Giessenbachstrasse 1, 85748 Garching, Germany.}
\item{School of Physics and Astronomy, University of Leicester, LE1 7RH, UK.}
\item{Institut d’Estudis Espacials de Catalunya (IEEC), Barcelona, Spain.}
\item{National Space Science Center, Chinese Academy of Sciences, Beijing, 100190, China.}
\item{Institute of Electrical Engineering, Chinese Academy of Sciences, Beijing, 100190, China.}

\end{affiliations}

\begin{abstract}
  
Massive stars end their lives as core-collapse supernovae, amongst which some extremes are broad-lined type Ic supernovae from Wolf-Rayet stars associated with long-duration gamma-ray bursts (LGRBs) having powerful relativistic jets. Their less-extreme brethren make unsuccessful jets that are choked inside the stars, appearing as X-ray flashes or low-luminosity GRBs. On the other hand, there exists a population of extragalactic fast X-ray transients (EFXTs) with timescales ranging from seconds to thousands of seconds, whose origins remain obscure. Here, we report the discovery of the bright X-ray transient EP240414a detected by the Einstein Probe (EP), which is associated with the type Ic supernova SN\,2024gsa at a redshift of 0.401. The X-ray emission evolution is characterised by a very soft energy spectrum peaking at $< 1.3$\,keV, which makes it different from known LGRBs, X-ray flashes, or low-luminosity GRBs. Follow-up observations at optical and radio bands revealed the existence of a weak relativistic jet that interacts with an extended shell surrounding the progenitor star. Located on the outskirts of a massive galaxy, this event reveals a new population of explosions of Wolf-Rayet stars characterised by a less powerful engine that drives a successful but weak jet, possibly owing to a progenitor star with a smaller core angular momentum than in traditional LGRB progenitors.
\end{abstract}

Extragalactic fast X-ray transients (EFXTs) are characterized by short-lived X-ray emissions originating from cosmological distances, with durations ranging from seconds to hours. Over 30 EFXTs have been detected in archival data from \textit{Chandra} and \textit{XMM-Newton}, yet their underlying nature remains elusive \cite{Jonker2013ApJ,Glennie2015,Xue2019,Lin2022,Quirola2022}, primarily due to insufficient timely follow-up observations. Proposed physical mechanisms to explain these events include the softer analogs of long gamma-ray bursts (LGRBs)\cite{Zhang2018pgrb.book, Liu2024arXiv, Yin2024arXiv}, shock breakouts of supernovae\cite{Klein1978ApJ, Soderberg2008Natur, Waxman2017}, unsuccessful jets\cite{Sakamoto2005,Campana2006}, or magnetar-powered X-ray emissions following binary neutron star mergers\cite{Zhang2013ApJ, Xue2019, Sun2023}, and other yet-to-be-explored astrophysical phenomena.

\begin{figure}[htbp]
    \centering
    \begin{minipage}[t][0.6cm][b]{0.48\textwidth}
        \centering
        \begin{minipage}[b]{\textwidth}
            \centering
            \begin{overpic}[width=0.82\textwidth]{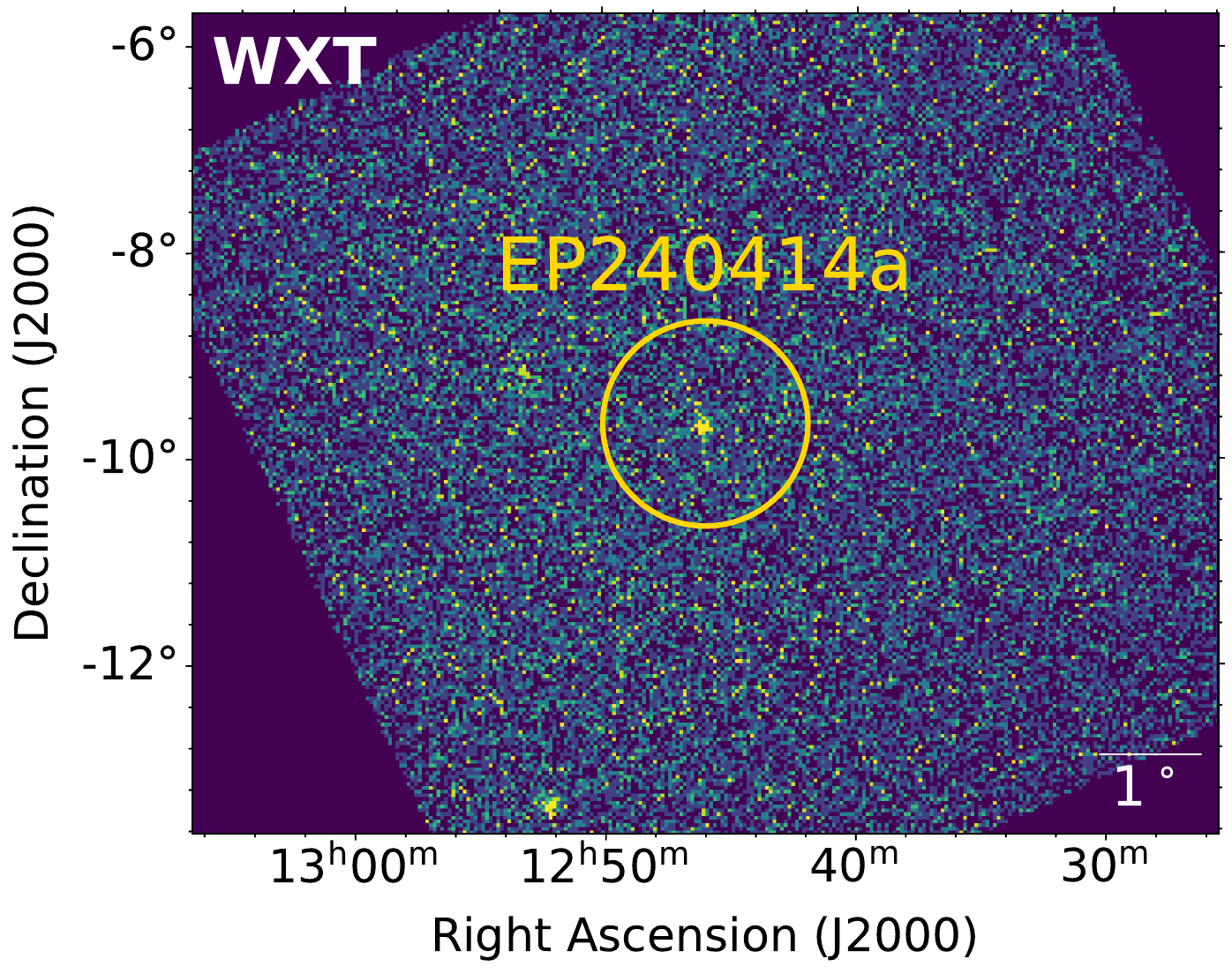}
            \put(-6,76){{\textbf{a}}}
            \end{overpic}
        \end{minipage}
        \begin{minipage}[b]{\textwidth}
            \centering
            \begin{overpic}[width=0.84\textwidth]{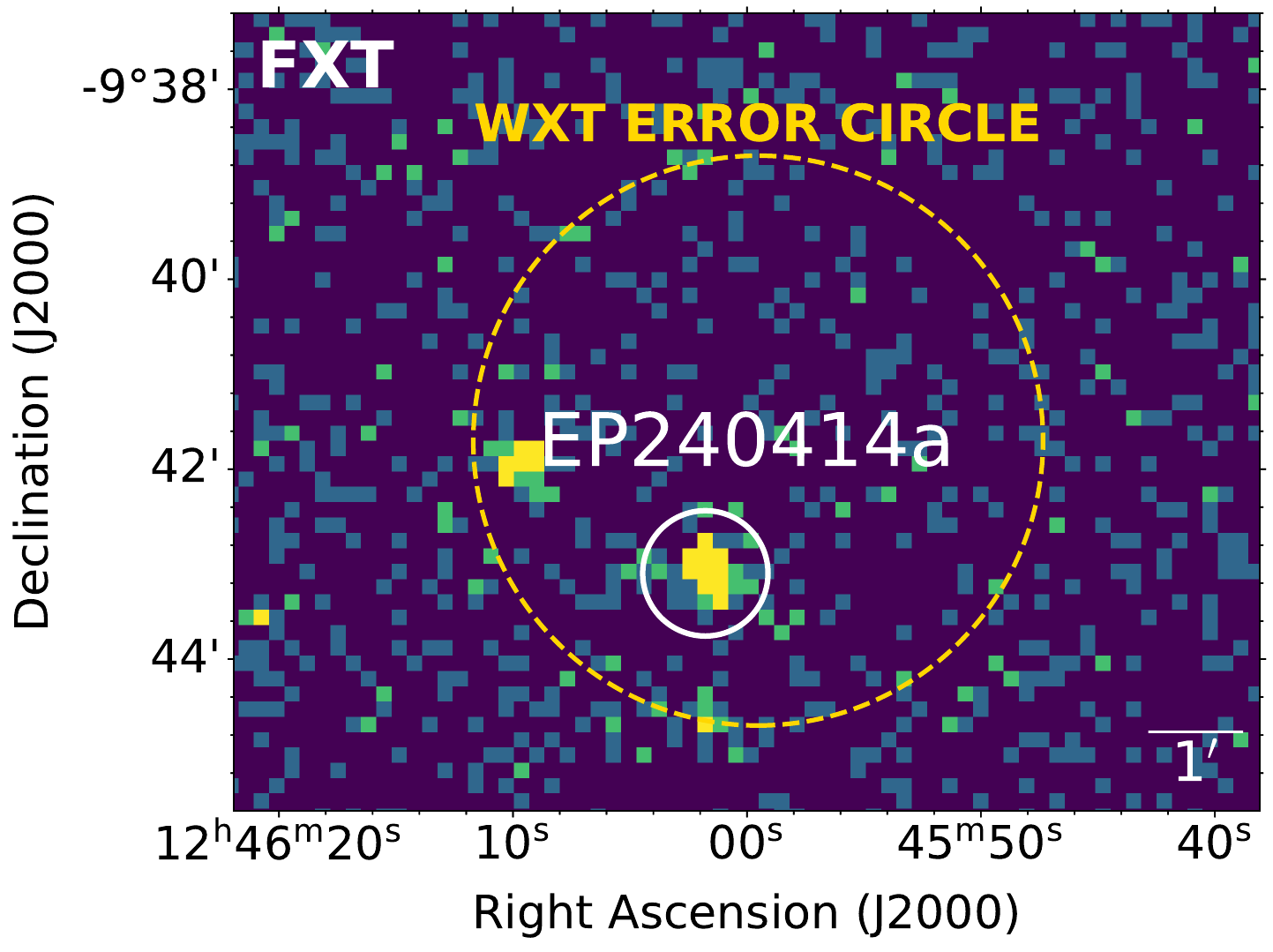}
            \put(-6,68){\textbf{b}}
            \end{overpic}
        \end{minipage}
    \end{minipage}
    \begin{minipage}[b]{0.5\textwidth}
        \centering
            \begin{overpic}[width=0.85\textwidth]{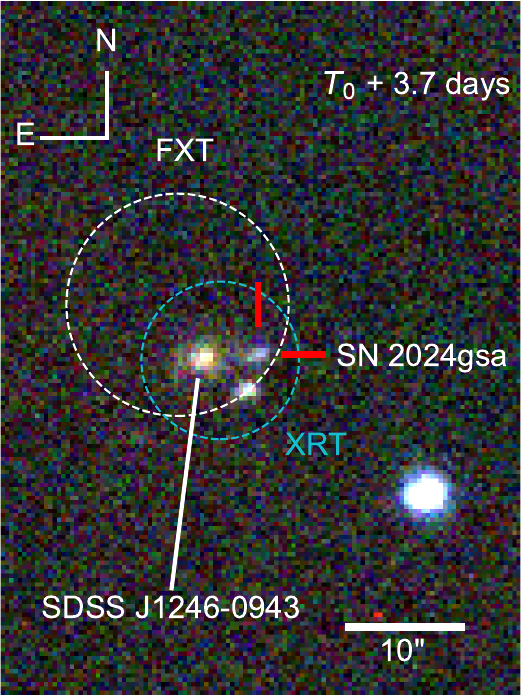}
            \put(-5,98){\textbf{c}}
            \end{overpic}
    \end{minipage}
    \begin{minipage}[t]{0.95\textwidth}
            \begin{overpic}[width=\textwidth]{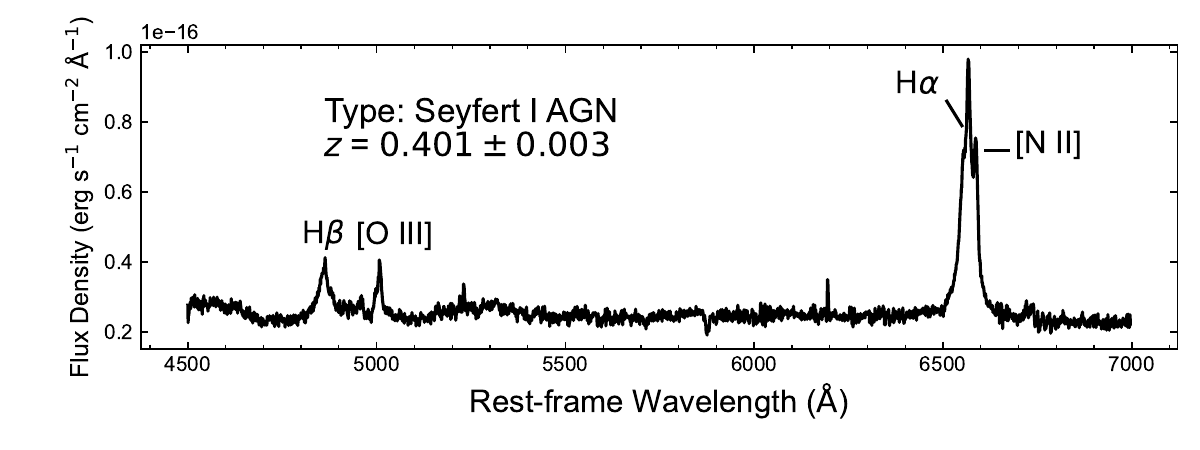}
            \put(1,34){\textcolor{black}{\textbf{d}}}
            \end{overpic}
    \end{minipage}
    \caption{\textbf{Multiwavelength images and host-galaxy spectrum of EP240414a/SN\,2024gsa.} \textbf{a}, The image of EP240414a in one of the WXT CMOS detectors. \textbf{b}, The image of EP240414a taken by FXT. The white circle represents the centre of FXT localisation with a radius of $40''$. The yellow circle shows the localisation of WXT with an error circle of $3'$. \textbf{c}, An SDSS $gri$-band composite image of SN\,2024gsa using LCO images (see Methods). SN\,2024gsa is marked with red ticks, while the white and cyan circles indicate the error circles of FXT and XRT, respectively. The host galaxy SDSS J1246-0943 is marked with a white tick. \textbf{d}, The spectrum of the nucleus of SDSS J1246-0943. Prominent emission lines, including H$\alpha$, H$\beta$, [O~III] $\lambda$5007, and [N~II] $\lambda$6583, are labeled. The spectral type of the galaxy and its redshift are also indicated.}
    \label{fig:three_images}
\end{figure}

EP240414a triggered the Wide-field X-ray Telescope (WXT) onboard the EP satellite in the 0.5--4\,keV band (Fig. \ref{fig:three_images}a) at \( T_0 = \) 09:49:10 on 14 April 2024\cite{2024GCN.36091....1L} (UTC).
Within a similar time window, no significant gamma-ray signals were detected in association with this event (see Methods). 
Subsequent follow-up observations of EP240414a revealed counterparts at soft X-ray (at $\sim$\( T_0~+\) 2\,hr, Fig. \ref{fig:three_images}b), optical (at $\sim$\( T_0~+\) 3\,hr\cite{Srivastav2024arXiv}, Fig. \ref{fig:three_images}c), and radio (at $\sim$\( T_0~+\) 9\,days\cite{2024GCN.36362....1B, Bright2024arXiv}) wavelengths.
Optical spectra suggested that it is likely associated with the galaxy SDSS~J124601.99-094309.3 (J1246) at a redshift of \( z = 0.401 \) (Fig. \ref{fig:three_images}d, Methods).

\begin{figure}
\centering
\begin{overpic}[width=0.40\textwidth]{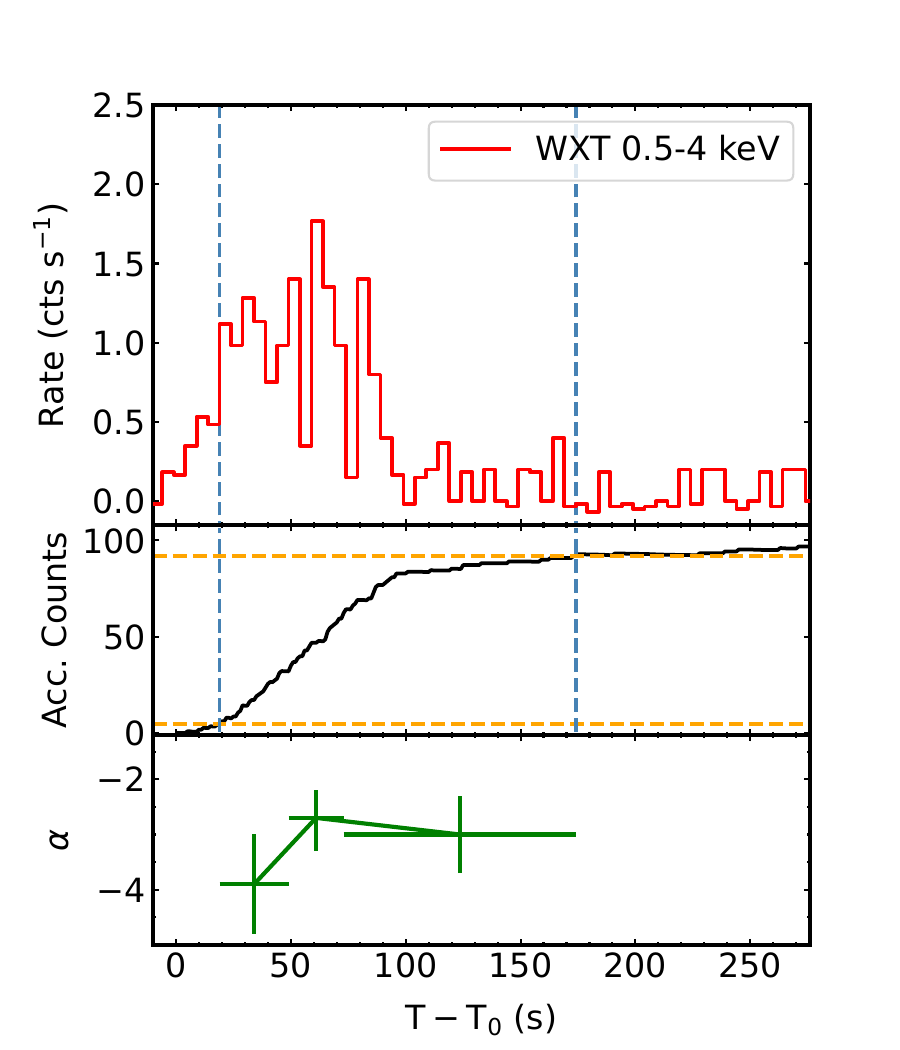}\put(1, 90){\bf a}\end{overpic} 
\begin{overpic}[width=0.40\textwidth]{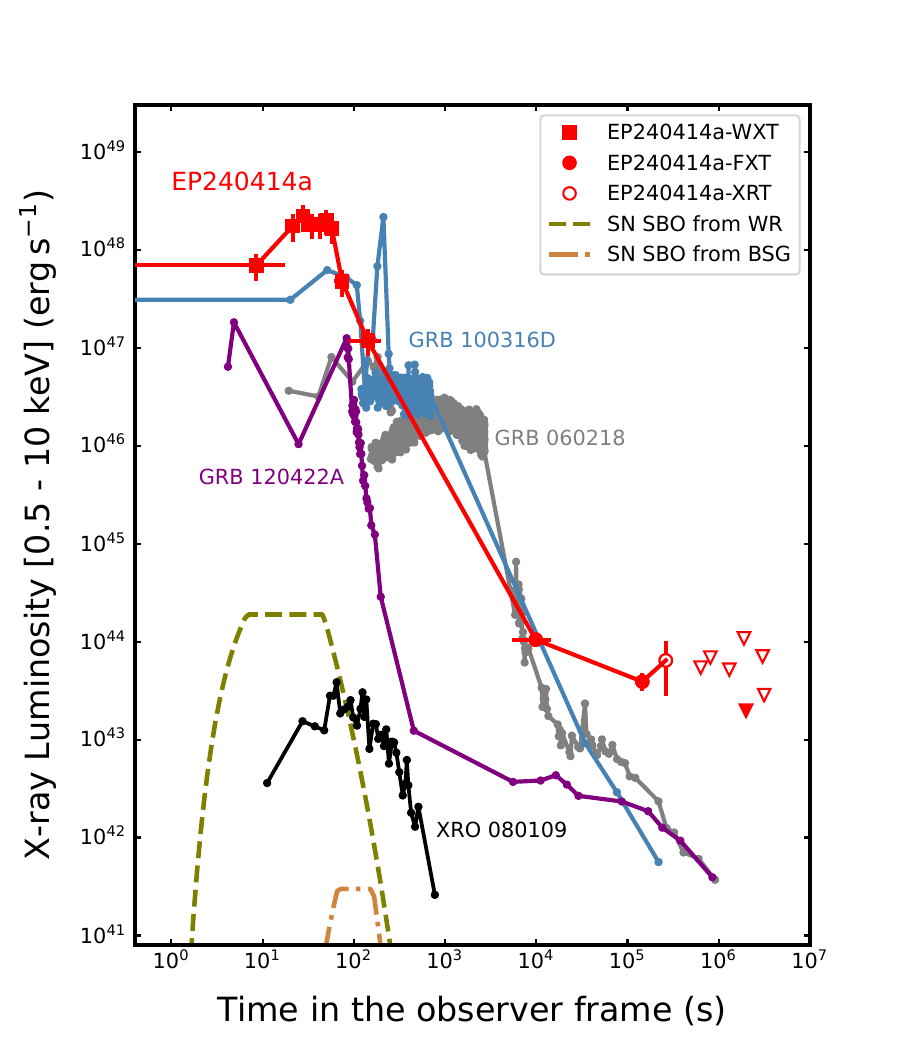}\put(1, 90){\bf b}\end{overpic} \\
\begin{overpic}[width=0.5\textwidth]{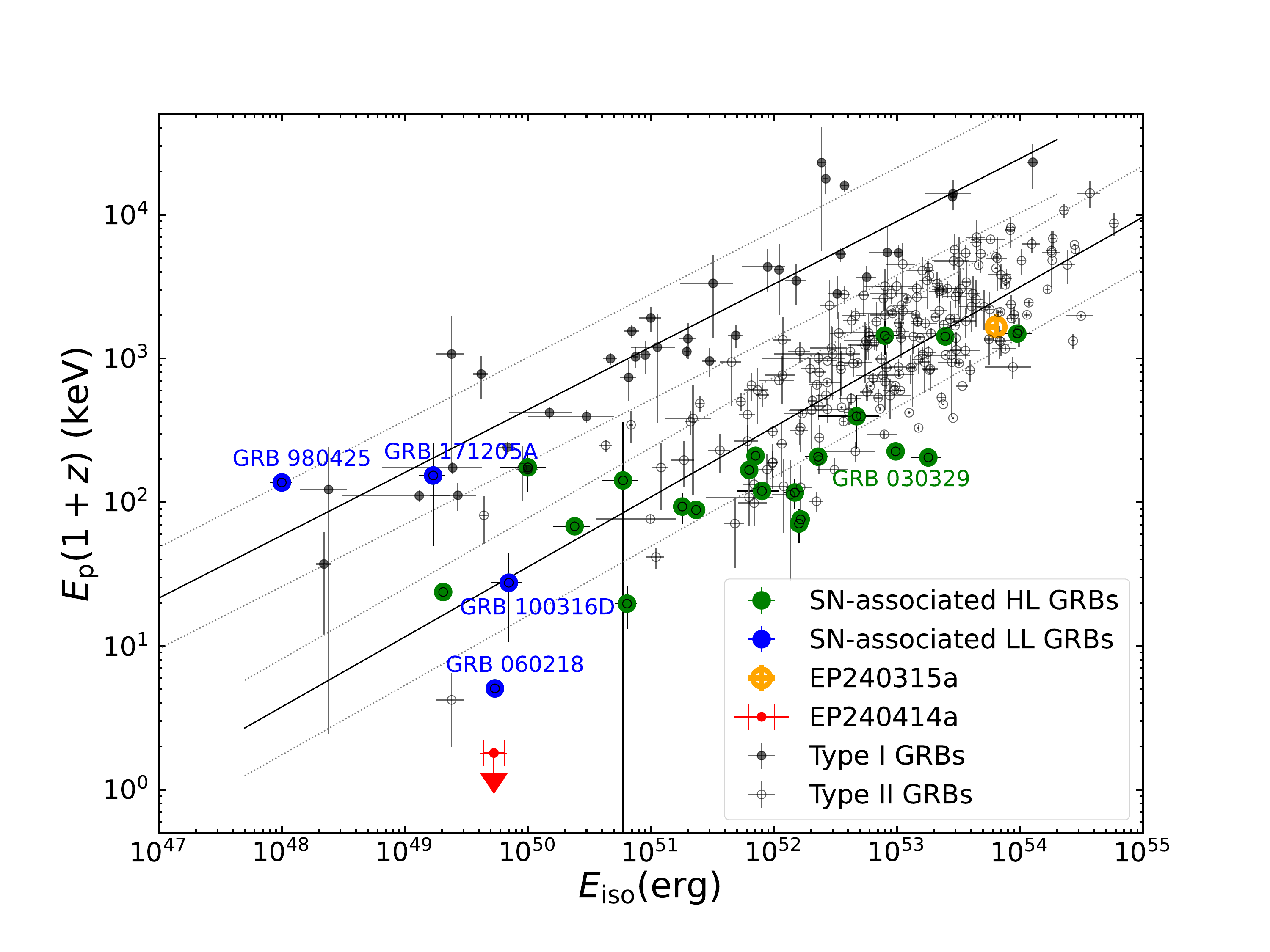}\put(3, 65){\bf c}\end{overpic} \\
\caption{\noindent\textbf{X-ray properties of EP240414a.} \textbf{a}, The light curve of the net count rate and the evolution of the photon index of EP240414a in the 0.5--4\,keV band. The bin size is 5\,s. The $T_{90}$ of the transient, during which 5\%--95\% of the fluence is observed, is 155\,s. Data are presented as the best-fit values with 1$\sigma$ confidence level. \textbf{b}, Long-term unabsorbed X-ray luminosity light curves in the 0.5--10\,keV band. The flux of the WXT data is derived with a count-to-flux conversion factor of $2.8 \times 10^{-9}$\,$\rm erg\,cm^{-2}\,ct^{-1}$ assuming the best-fit result for the average spectrum fitted with the absorbed power-law model in Extended Data Table \ref{tab:spectral_analysis}. The dashed and dash-dotted lines respectively represent the X-ray luminosity of SN shock breakout from Wolf-Rayet (WR) star and blue supergiant (BSG) explosions\cite{Nakar2010ApJ}. The data of low-luminosity GRBs 060218, 100316D, and 120422A are obtained from BAT and XRT observations in the Swift Burst Analyser \cite{Evans2010}.
The data of XRO\,080109 are obtained from Ref. \cite{Modjaz2009} \textbf{c}, The rest-frame peak energy versus isotropic energy correlation (Amati relation). 
The burst is an outlier compared with classical GRBs and some low-luminosity GRBs. Type I and Type II GRBs refer to physically distinct GRB types with a compact merger origin and with a massive star core collapse origin, respectively. The isotropic energy for EP240414a is measured in the range of 0.5--4 keV, which dominates that of 1--$10^4$ keV due to its soft photon index. The peak energy of GRB 060218 has a wide range\cite{Kaneko2007ApJ}, and the value obtained from the average of the spectral fits is plotted. The solid lines represent the best-fit correlations and the dotted lines show the 1$\sigma$ confidence region. Uncertainties are given at the 1$\sigma$ confidence level.}
\label{fig:wxt_lc_spec}
\end{figure}

The X-ray burst captured by the WXT is characterised by a single pulse with marginal variability, as shown in Fig. \ref{fig:wxt_lc_spec}a. The $T_{90}$ of the transient, i.e. the time during which the central 90\% of the fluence is observed, is 155\,s (Table \ref{tab:obs_prob}). The integrated spectrum within $T_{90}$ can be fitted by an absorbed power-law model with a photon index $\alpha = -3.1_{-0.8}^{+0.7}$, and an intrinsic absorption of $N_{\rm int} = 7.4^{+4.1}_{-3.7} \times 10^{21}$\,$\rm cm^{-2}$ in excess of the absorption within the Milky Way Galaxy (see Methods and Extended Data Figure \ref{fig:wxt_spec}). The soft spectrum fitted with the power-law model indicates that the energy peak $E_{\rm peak}$ is near or below 
the lower limit of WXT's energy range. 
The spectral fitting result with the absorbed broken power-law model implies an upper limit of less than $1.3$ \,keV for the $E_{\rm peak}$ (see Methods and Extended Data Table \ref{tab:spectral_analysis}). 
This is considerably lower than that of the X-ray transients detected by the Einstein Probe in relation to the GRBs, such as EP240315a\cite{Liu2024arXiv}.
At the redshift of 0.401, the absorbed peak isotropic X-ray luminosity in the 0.5--4\,keV band is $1.3 \times 10^{48}$\,$\rm erg\,s^{-1}$, which exceeds that of the supernova (SN) shock-breakout event XRO\,080109 by more than four orders of magnitude\cite{Soderberg2008Natur}. As shown in Fig. \ref{fig:wxt_lc_spec}b, this luminosity is also much higher than the predicted value for nonrelativistic SN shock-breakout models\cite{Nakar2010ApJ}. The luminosity of EP240414a falls into the range of low-luminosity GRBs\cite{Virgili2009MNRAS, Hjorth2012grb, Sun2015ApJ} which have also been proposed to be relativistic shock breakouts\cite{Wang_2007,Nakar2012} . This transient, however, is much softer than the prediction based on the so-called closure relation of the relativistic shock-breakout model\cite{Nakar2012} given the observed duration and isotropic energy. The X-ray fluence indicates a total isotropic equivalent energy of $5.3_{-0.6}^{+0.8}\times 10^{49}$\,$\rm erg$ measured in the range of 0.5--4\,keV. Combining the upper limit of the peak energy, we find that this event is a unique outlier in the so-called ``Amati relation,'' which was believed to be generally satisfied by LGRBs, X-ray flashes, or low-luminosity GRBs\cite{Amati2002,Sakamoto2005,Zhang2009} (see Fig. \ref{fig:wxt_lc_spec}c). The soft X-ray spectrum, the absence of a gamma-ray counterpart, a significant outlier in the Amati relation and a large offset from its host galaxy (see below) hint that EP240414a likely have a different origin from previously known bursts.

Motivated by the discovery of EP240414a, we initiated multiwavelength follow-up observations at X-ray, optical, near-infrared (NIR), and radio bands, extending up to \( T_0~+\) 50\,days (see Methods and Extended Data Tables \ref{tab:FXT_XRT_obsinfo} and \ref{table:opt_ir}). An X-ray counterpart was detected by the EP Follow-up X-ray Telescope (FXT) $\sim 2$\,hr after the trigger\cite{2024GCN.36129....1G} with an absorbed flux of $1.8 \times 10^{-13}$\,$\rm erg\,cm^{-2}\,s^{-1}$ in the 0.5--10\,keV band (see Methods), implying a rapid decline in its X-ray flux by more than three orders of magnitude within about 3\,hr after the detection. The spectra observed in the two FXT observations can be fitted with an absorbed power-law model with a photon index of $-2.2^{+0.3}_{-0.4}$ (Methods). During the phase from $\sim T_0+2$\,hr to the detection by the {\it Swift} X-ray Telescope at around $\sim T_0+4$\,days, the evolution of the X-ray emission enters a plateau before further decaying.

\begin{figure}
\centering
\begin{tabular}{c}
\includegraphics[width=0.95\textwidth]{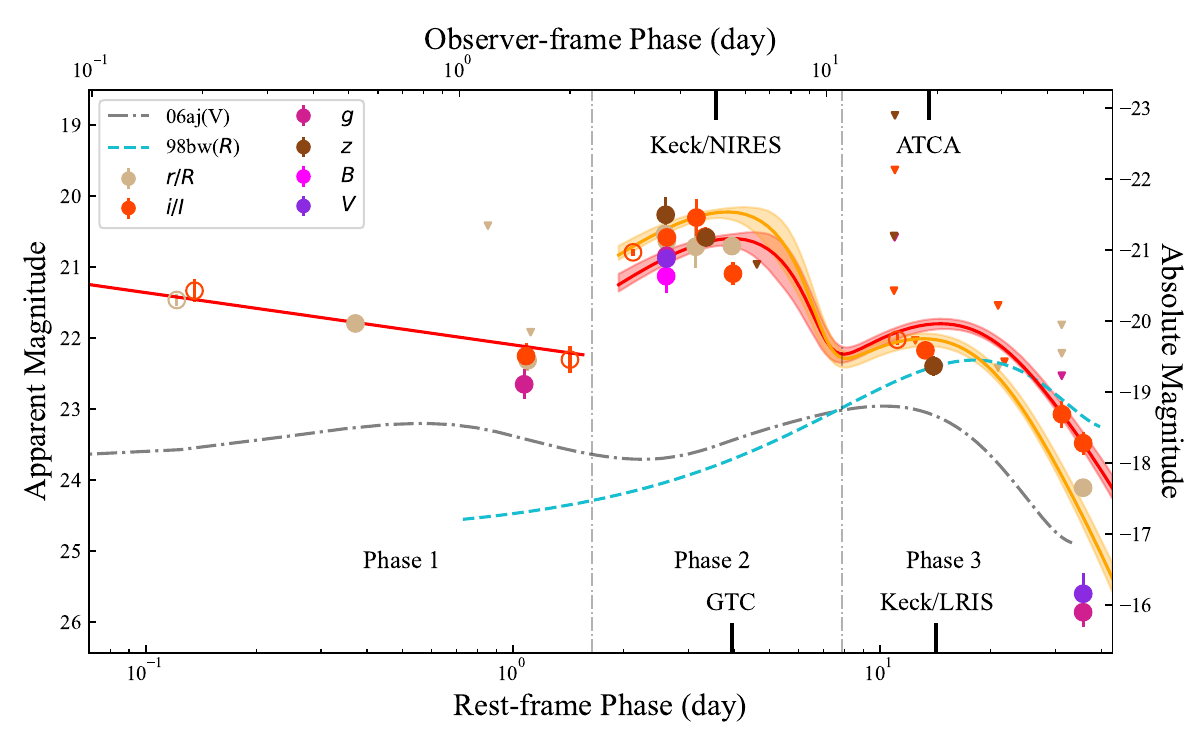}
\end{tabular}
\caption{\noindent\textbf{Optical and NIR light curves of SN\,2024gsa.} 
Magnitude is plotted against the rest-frame phase on the lower abscissa and the observer-frame phase on the upper abscissa. The left ordinate denotes apparent magnitude, while the right ordinate is absolute magnitude. Photometric data points represent median measurements, with error bars indicating 1$\sigma$ uncertainties. Data from different bands are distinguished using various colours. The \textit{griz}-band photometry is given in the AB magnitude system, while all other bands are in the Vega magnitude system. Circles represent data points, whereas triangles give 3$\sigma$ upper limits. All magnitudes shown have been corrected for Galactic extinction. The dashed line corresponds to the extinction-corrected light curves of SNe\,1998bw and 2006aj placed at the same distance as SN\,2024gsa for comparison\cite{Clocchiatti2011,Campana2006}. The light curve is segmented into three phases by two vertical dot-dashed lines (see the main text for details). Vertical solid lines mark the epochs of spectroscopy from the GTC and Keck~I/II telescopes, as well as radio observations from ATCA. Data points represented by open circles are from literature \cite{Srivastav2024arXiv,vanDalen2024arXiv}. The modeled light curves for the $i$ band (red solid line) and $r$ band (orange solid line) are also displayed. Shaded regions indicate the 1$\sigma$ standard deviation around the mean fitted magnitude in Phases 2 and 3.
}
\label{fig:lightcurve}
\end{figure}

Fig. \ref{fig:lightcurve} shows the overall luminosity evolution of the optical counterpart SN\,2024gsa collected within $\sim 50$\,days after the discovery (Methods); the photometry we obtained is shown in Extended Data Table \ref{table:opt_ir}. Optical emission associated with EP240414a was detected within a few hours after the X-ray trigger, possibly representing the afterglow of the prompt X-ray emission (Phase 1).
At \(T_0 + 4\)\,days, a second optical bump was observed peaking at $\sim -21.5$\,mag in the $z$ band, which represents the most luminous non-afterglow optical emission ever recorded for an EFXT (Phase 2). The third optical bump appears at \(\sim T_0 + 15\)\,days, fainter but apparently broader than the second one, with a peak luminosity of $M_{I} \approx -19.5$\,mag (Phase 3).  
Follow-up radio observations reveal that the radio counterpart, detected at \(\sim T_0 + 19\)\,days, exhibited a rising spectrum between 5.5 and 9\,GHz (Methods, Extended Data Figure \ref{fig:afterglow_fit} and Extended Data Table \ref{table:flux_density}), consistent with synchrotron emission from a relativistic jet. The radio luminosity is comparable to that of bright GRBs, such as GRB\,030329, and is significantly brighter than that of most SNe (Extended Data Figure \ref{fig:radio}).

\begin{figure}
\centering
\begin{tabular}{c}
\begin{overpic}[width=0.9\textwidth]{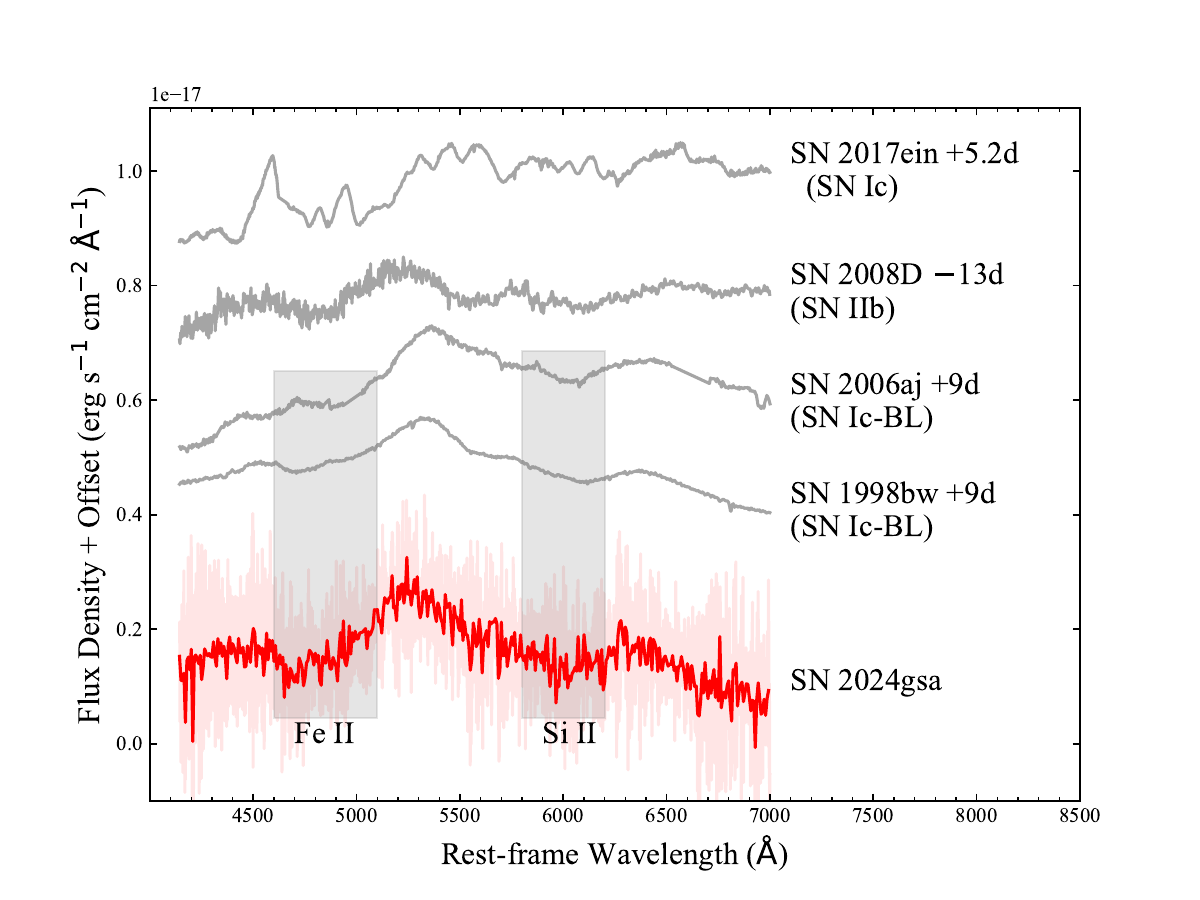}\end{overpic} \\
\end{tabular}
\caption{\noindent\textbf{
Comparison of the optical spectrum of SN\,2024gsa with those of selected stripped-envelope SNe.} The spectrum of SN\,2024gsa (light red), taken on \(T_0 + 20\)\,days, is compared with those of four selected SNe~Ib/c (grey), including
two SNe~Ic-BL, SN\,1998bw\cite{Patat2001} and SN\,2006aj\cite{Pian2006Nature}; a normal SN~Ic, SN\,2017ein\cite{Xiang2019}; and an SN associated with an X-ray outburst (XRO), SN\,2008D\cite{Mazzali2008}. A rebinned version of the spectrum (red), generated using a bin width of 20 \AA, is also overplotted. The rest-frame phases, relative to the time of \( B \)-band or \( V \)-band maximum brightness, are labeled next to the corresponding spectra. The wavelengths are redshift-corrected to the rest frame. Grey regions indicate two commonly seen broad absorption features in SNe~Ic-BL: Fe~II blended lines and the Si~II $\lambda$6355 line.}
\label{fig:keck}
\end{figure}

To identify the properties of the optical counterpart SN\,2024gsa, three optical spectra were obtained (Fig. \ref{fig:keck} and Extended Data Figure \ref{fig:GTC}). The redshift-corrected spectrum taken with LRIS on the Keck~I telescope\cite{Oke1995} at $\sim$ \(T_0 + 20\) days (Fig. \ref{fig:keck}) is consistent with a broad-lined Type Ic SN (SN~Ic-BL) at an age of $\sim 10$\,days after the explosion, with a best-fit redshift of 0.38$\pm$0.02. 
The overall broad spectral profile of SN\,2024gsa is quite similar to that of other GRB-associated SNe (GRB-SNe)\cite{WoosleyBloom2006,Cano2017}, such as SN\,1998bw associated with GRB\,980425 \cite{Galama1998Natur} and SN\,2006aj associated with GRB\,060218 \cite{Pian2006Nature}, with the broad absorption features at 460--490\,nm and 590--620\,nm attributed to Fe~II blended lines and Si~II $\lambda$6355, respectively. Furthermore, the absence of hydrogen and helium absorption features in the spectrum indicates that the progenitor star had lost its outer H and He envelopes prior to the explosion. Such characteristics align with a stripped-envelope progenitor, typically associated with Wolf-Rayet stars \cite{Crowther2007}. Combining these spectral features with an ejecta velocity of about 16,700\,km\,s$^{-1}$ derived from the Si~II absorption feature (Methods), we infer that SN\,2024gsa originates from an energetic explosion of a Wolf-Rayet star.

A luminous spiral galaxy, J1246, is observed in close proximity to SN\,2024gsa; its redshift is consistent with that of SN\,2024gsa (Methods), suggesting it is the host galaxy of EP240414a/ SN\,2024gsa. A spectrum of J1246 further revealed the presence of an active galactic nucleus (AGN) at its core (Methods). The projected offset of \(26.3 \pm 0.1\)\,kpc from the galaxy centre is unusually large for SNe~Ic-BL\cite{Japelj2018} (Extended Data Figure \ref{fig:offset}). Such a host galaxy and location in a galaxy is quite different from those of LGRBs and low-luminosity GRBs, which are typically dwarf star-forming galaxies with the bursts occurring in the brightest region in the galaxies\cite{fruchter2006}. On the other hand, the spectrum of J1246 indicates that the central region of the galaxy has low metallicity. The SN, located on the outskirts of J1246, is likely in a region with even lower metallicity than the galaxy’s nucleus. Thus, the possibility of an association with an undetected satellite dwarf galaxy or a distant star-forming H~II region cannot be ruled out (see Methods).

Multiband observations suggest that EP240414a/SN\,2024gsa originated from the core-collapse explosion of an extragalactic massive star with its envelope significantly stripped prior to the explosion. The core collapse induced shock waves sequentially moving through the stellar envelope and a previously ejected circumstellar medium (CSM), ultimately producing an optical bump on a timescale of days, rather than an X-ray signal on a shorter timescale. Detailed modeling shows that as long as the radius and mass of the ejected shell reach $ R_{\rm ext} 
\approx 2.41 \times 10^{14}$\,cm and $ M_{\rm ext} \approx 0.33\,M_{\odot} $, the analytical shock-cooling emission from this extended material \cite{Margalit2022} can well fit the optical light curve of SN\,2024gsa during Phase 2 (see Methods). 
The optical light curve in Phase 3 is powered by the radioactive decay of
$^{56}$Ni, typical of SNe Ic-BL. We fit the multi-band optical light curves
using a widely adopted semi-analytic model for supernova light curve modeling\cite{Arnett1982}. We derive the following explosion parameters from the fit: ejecta mass
$M_{\text{ej}} \approx 2.38 M_{\odot}$ and synthesized nickel mass
$M_{\text{Ni}} \approx 0.74 M_{\odot}$, which exceeds the nickel
mass reported for SN\,1998bw ($0.3 - 0.6  M_{\odot}$)\cite{Cano2017}.
This is consistent with the higher luminosity observed for this event. Additionally, the derived ejecta velocity \(v_{\text{ej}} \approx 1.5 \times 10^9 \, \text{cm s}^{-1}\) matches that measured from the observed SN spectrum.

Excluding the radiation signals originating from the Type Ic SN and its breakout from a dense shell, other multiwavelength observations reveal the emission signature of a weak, relativistic jet.  
The achromatic decay in the X-ray and optical data up to \( T_0~+ \) 2 days is consistent with the standard afterglow synchrotron emission from a decelerating relativistic jet, with a lower limit on the initial Lorentz factor (\( \Gamma_0 \)) of $\sim 13$. The isotropic kinetic energy of the jet is estimated to be \(\sim~10^{51}\) erg with a CSM of density \( \sim 1 \, \text{cm}^{-3}\). The late-time radio afterglow is also consistent with the afterglow model but requires a lower CSM density (see Methods).

The unique properties of EP240414a, including its extremely soft spectrum that does not agree with any previously known LGRBs and relatives, a substantial amount of pre-explosion ejected mass, its peculiar host galaxy, and the unusual position within it, suggest that it likely represents the prototype of a previously unknown population of envelope-stripped Wolf-Rayet stars. EP240414a/SN\,2024gsa-like events bridge the gap between traditional GRBs and those broad-lined SNe~Ic that do not have any high-energy counterparts, suggesting a diverse zoo of progenitor stars. The lower limit of the intrinsic event rate density for EFXTs similar to EP240414a, derived from the eight months of operation of EP-WXT, is $\approx 0.3$ $\rm Gpc^{-3}yr^{-1}$ (see Methods), suggesting that this population contributes to a non-negligible fraction of cosmic explosions. The peculiar properties of EP240414a may be related to moderately high core angular momentum, lower magnetic field, or higher mass and/or metallicity of the progenitor star. Future studies of EFXTs analogous to EP240414a, utilising facilities such as the EP, will offer valuable insights into the physical processes governing the death of massive stars, including mass loss, angular momentum transfer, jet formation, energy conversion between the jet and the stellar envelope, and the mechanisms driving the initial kinetic energy of SNe. 

\clearpage

\begin{table*}
\centering
\begin{threeparttable}
\caption{\textbf{Properties of EP240414a.} Errors represent the 1$\sigma$ uncertainties.}
\label{tab:obs_prob}
\begin{tabular}{lc}
\toprule
Observed Properties &  EP240414a\tnote{*} \\
\hline
\textbf{Soft X-Ray [0.5--4\,keV]} & \\
\hline
Duration $T_{90}$ (s) & $155^{+64}_{-22}$ \\
Photon Index $\alpha$ & $-3.1^{+0.7}_{-0.8}$  \\
Intrinsic Absorption $N_{\rm int}$ ($\rm cm^{-2}$ ) &  $7.4^{+4.1}_{-3.7} \times 10^{21}$\\
Peak Energy ($\rm keV$) & $< 1.3$\\
Peak Flux ($\rm erg\,cm^{-2}\,s^{-1}$) & $(2.2 \pm 0.7) \times10^{-9} $  \\
Peak Luminosity ($\rm erg\,s^{-1}$) & $(1.3\pm 0.4)\times10^{48} $ \\
Total Fluence ($\rm erg\,cm^{-2}$) & $1.2^{+0.3}_{-0.2} \times10^{-7} $ \\
Isotropic Energy ($\rm erg $) & $5.3_{-0.9}^{+1.2}\times10^{49} $\\
\hline
\textbf{Host Galaxy:} & \\
\hline
Redshift & $0.401 \pm 0.003$ \\
Projected Offset ($\rm kpc$) &   \(26.3 \pm 0.1\)\, \\
\hline
\textbf{Associations:} & \\
\hline
Gamma-ray Counterpart & No \\
Supernova & SN\,2024gsa (Ic-BL)  \\
\bottomrule
\end{tabular}
\begin{tablenotes}
\footnotesize
\item[*] The source position is R.A. = $12^h46^m01.682^s \pm 0.003^s$, Dec. = $-09^\circ43'08.13'' \pm 0.33''$, inferred from the radio observation.
\end{tablenotes}
\end{threeparttable}
\end{table*}

\clearpage

\clearpage

\clearpage

\clearpage

\clearpage

\section*{Methods}
\subsection{Observations and data reduction.\\} \label{sec:obs}

\noindent\textbf{X-rays.} The Einstein Probe (EP)
\cite{Yuan2022}, launched on 9 Jan. 2024, is a mission led by the Chinese Academy of Sciences (CAS), in collaboration with the European Space Agency (ESA) and the Max Planck Institute for Extraterrestrial Physics (MPE), Germany, dedicated to time-domain high-energy astrophysics. The Wide-field X-ray Telescope (WXT) is one of the two main payloads on the EP. It employs novel lobster-eye micro-pore optics (MPO) to enable a large instantaneous field of view (FoV) of 3600 square degrees and a sensitivity of $\sim 2.6\times10^{-11}$\,$\rm erg\,cm^{-2}\,s^{-1}$ in the 0.5--4\,keV band for an exposure time of 1\,ks. 

EP240414a was detected by the WXT during the commissioning phase of EP at 09:49:10 on 2024-04-14 UTC. It also triggered the WXT on-board processing unit at 2024-04-14T09:50:12 with a signal-to-noise ratio of 8.7. The WXT on-board trigger requires the signal-to-noise ratio above 8. The X-ray photons were processed using the data reduction pipeline and calibration database (CALDB) designed for WXT (Liu et al. in prep.). The CALDB was built combining the results of the on-ground calibration experiments and in-orbit calibration observations (Cheng et al. in prep.). The overall procedure is similar to that applied to the pathfinder of WXT, i.e. {\em Lobster Eye Imager for Astronomy}\cite{zhang2022ApJ,Cheng2024}. The positions of photons were re-projected to celestial coordinates (J2000). The Pulse Invariant value (the energy in the channel unit) of each event was calculated according to the bias and gain stored in CALDB. After flagging bad/flaring pixels and assigning grade, the events with grade 0--12 and without anomalous flags were selected to form the cleaned event file and the image in the 0.5--4\,keV band for further source detection. The light curve and the spectrum of the source and background in a given time interval were extracted from a circle with a radius of 9 arcmin and an annulus with inner and outer radii of 18 arcmin and 36 arcmin, respectively.

Follow-up observations were carried out with the EP Follow-up X-ray Telescope (FXT)\cite{Chen2020SPIE} and {\it Swift} X-Ray Telescope (Extended Data Table \ref{tab:FXT_XRT_obsinfo}), spanning from 2\,hr to 50\,days after the trigger. The FXT is one of the main payloads of EP in the 0.3--10\,keV range. It consists of two co-aligned modules (FXT-A and FXT-B), each containing 54 nested Wolter-I paraboloid-hyperboloid mirror shells. Both the source and background regions of the FXT data were processed using the FXT data analysis software (fxtsoftware\;v1.10; http://epfxt.ihep.ac.cn/analysis). The process involved particle event identification, PI conversion, grade calculation and selection (grade 0--12), bad and hot pixel flag and selection of good time intervals using housekeeping files. The pipeline finally resulted in the cleaned event files, energy spectra and response files.

\noindent\textbf{Gamma rays.} 
The {\it Fermi} Gamma-ray Burst Monitor was temporarily turned off owing to the South Atlantic Anomaly during the trigger time. 
{\it Konus-Wind} was observing the whole sky around 
EP240414a. No significant signal was observed during $T_0 \pm 2000$ s. The upper limit (90\% confidence level) of the flux in the 15--350\,keV band of Konus-Wind is $2.7 \times 10^{-7}$ $\rm erg\,cm^{-2}\,s^{-1}$ (2.944
s scale) assuming a power-law model with $\alpha=-3.1$. The {\it Swift} Burst Alert Telescope (BAT) did not detect any significant signal at $T_0 \pm 50$\,s. The direction of the target was slightly outside the coded FoV of BAT. The upper limits of BAT in the 15--350\,keV band for a power-law model with $\alpha=-3.1$ are 
$1.1 \times 10^{-6}$ $\rm erg\,cm^{-2}\,s^{-1}$ for a 1\,s time scale and $3.5 \times 10^{-7}$ $\rm erg\,cm^{-2}\,s^{-1}$ for a 16\,s time scale. 

\noindent\textbf{Optical and near-infrared photometric observations.\\} 
Optical photometric observations were carried out using numerous facilities. 
Through the Global Supernova Project\cite{2017AAS...23031803H}, \textit{BVgri}-band images were obtained using the network of 1.0\,m telescopes of the Las Cumbres Observatory (LCO)\cite{Brown2013} at Siding Spring Observatory (New South Wales, Australia), South African Astronomical Observatory (Sutherland, South Africa), and Cerro Tololo Inter-American Observatory (Cerro Tololo, Chile). Additionally, imaging observations were conducted with  the Nordic Optical Telescope (NOT), the Palomar 5\,m telescope, the Gran Telescopio Canarias (GTC) (GTCMULTIPLE2G-24A; PI: Nancy Elias-Rosa), the Keck~I telescope, the Xinglong 2.16\,m Telescope (XLT), the Lijiang 2.4\,m Telescope (LJT), the Southern Astrophysical Research Telescope (SOAR) in Chile, the AZT-22 telescope in Uzbekistan, the VLT Survey Telescope (VST) in Chile, the Aletai Telescopes (ALT), and the Burst Observer and Optical Transient Exploring System\cite{Castro2023} (BOOTES) telescope (BOOTES-4/MET).

To enhance the signal-to-noise ratio (S/N), image stacking was performed using the \texttt{REPROJECT} package (\url{https://reproject.readthedocs.io/en/stable/index.html}). Point-spread function (PSF) photometry for images from LCO, GTC, SOAR, VST, AZT-22, ALT, BOOTES-4, and XLT was carried out using \texttt{AutoPhOT} (\url{https://github.com/Astro-Sean/autophot/})\cite{AUTOPHOT}. 
\textit{BVR}- and \textit{griz}-band PSF flux was calibrated in Vega \cite{1992AJ....104..340L} and AB \cite{2017ApJS..233...25A} magnitude systems, respectively.
For optical images from NOT, due to the large seeing and contamination from J1246, the PSF did not provide a good fit. Therefore, aperture photometry was performed on the stacked frames, and the flux was calibrated against neighboring reference stars from the Pan-STARRS (PS1) field \cite{2016arXiv161205560C} in the AB magnitude system.

Besides the optical images, we also collected the NIR images of SN\,2024gsa using the Wide-field Infrared Camera (WIRC\cite{Wilson2003}) on the Palomar 5\,m telescope (P200). For WIRC data, a Python-based pipeline applied standard reduction techniques, stacked using Swarp\cite{Bertin2002} and photometrically calibrated against the 2MASS point-source catalog in the Vega magnitude sytem\cite{Skrutskie2006}. Image subtraction was performed following the ZOGY algorithm\cite{Zackay_2016}, using its implementation in Python\cite{Guevel2017Pyzogy}.

Very deep images were obtained with the Keck-I telescope and the Low Resolution Imaging Spectrometer (LRIS)\cite{1995PASP..107..375O} at the W. M. Keck Observatory on June 3, at a phase of about 50 days after the EP trigger. The data were processed using the \textsc{LPipe} data-reduction pipeline\cite{2019PASP..131h4503P}. SN\,2024gsa was detected in $V$-, $g$-, $R$-, and $I$-band images. PSF photometry was performed on all images using AutoPhOT. For the flux calibration, photometric data from Gaia DR3\cite{2016A&A...595A...1G,2023A&A...674A...1G} were transformed into the Johnson-Cousins photometric system ($BRI$ magnitudes) by referencing standard stars within the field of view, following the Gaia DR3 documentation\cite{2022gdr3.reptE....V}. The $VRI$-band flux was calibrated using these reference stars in the Vega magnitude system, while the \textit{g}-band flux was calibrated against reference stars from the PS1 field in the AB magnitude system.

The final optical and NIR photometric results are summarized in Extended Data Table \ref{table:opt_ir}, and the corresponding light curves are shown in Figure \ref{fig:lightcurve}.

\noindent\textbf{Optical and near-infrared spectroscopic observations.\\} 
Three spectra were collected for SN\,2024gsa. On April 19.9 (MJD = 60419.9; \(T_0~+\)\,5.5 days), we took a spectrum with the OSIRIS+ mounted on the GTC. The spectrum underwent bias and flat-field correction, extraction, and wavelength and flux calibration using the {\sc FOSCGUI}
pipeline (\url{http://sngroup.oapd.inaf.it/foscgui.html}), adapted for the OSIRIS+ instrument. Telluric lines were also removed from the spectra. 
The blue part of Extended Data Figure \ref{fig:GTC} presents the redshift‐corrected (see below for redshift discussion) GTC/OSIRIS+ spectrum of SN\,2024gsa. To better illustrate the spectrum given the high noise level in the original data, a rebinned spectrum using a bin width of 20 \AA\ is also overplotted. The spectrum exhibits a featureless continuum. However, this spectrum does not provide sufficient information to determine the physical properties of the transient.

A NIR spectrum of SN\,2024gsa was obtained using NIRES\cite{2004SPIE.5492.1295W} on the Keck~II telescope at a similar phase (i.e., on April 19). We reduced the NIRES data using PypeIt (\url{https://pypeit.readthedocs.io/en/release/index.html\#pypeit-bibtex-entries}),
which performs flat-field calibration, background subtraction, and spectral extraction. Flux calibration, coaddition, and telluric correction are done using a sensitivity function fitted to the telluric-star observation. 
The red part of Extended Data Figure\ref{fig:GTC} presents the redshift‐corrected (see below for redshift discussion) Keck/NIRES spectrum of SN\,2024gsa. A rebinned version of this spectrum, using the same binning procedure as applied to the GTC spectrum, is also overplotted. Owing to the insufficient S/N, the accuracy of the flux calibration is limited to within a factor of a few.

On May 4.3 (\(T_0 +20\)\,days), we obtained a long-slit spectroscopy of SN\,2024gsa using Keck-I/LRIS, which is equipped with an atmospheric dispersion corrector. The data were processed with the \textsc{LPipe} data-reduction pipeline. The spectra of SN\,2024gsa and host J1246, a potential host galaxy of SN\,2024gsa source located 5.7$^{\prime\prime}$ away (see Fig. \ref{fig:three_images} (a)), were obtained simultaneously. By template matching using GELATO\cite{2008A&A...488..383H} (gelato.tng.iac.es) and Superfit\cite{2005ApJ...634.1190H}, we classified SN\,2024gsa as an SN~Ic-BL with redshift $z = 0.38\pm0.02$. It is worth noting that the spectrum of SN\,2024gsa was obtained before its optical maximum brightness (Fig. \ref{fig:lightcurve}), which is earlier than the phase of the best-fit spectrum from both GELATO and Superfit. 
Given the slow spectral evolution of SNe~Ic-BL over several days, we conclude that the difference does not affect the classification of SN\,2024gsa. The relatively large uncertainty in the redshift is due to the lack of narrow emission lines and the low S/N of the spectrum. The redshift-corrected spectrum displays a continuum with two broad spectral features: a blend of Fe~II lines and a separate Si~II $\lambda$6355 feature. We measured the Si~II $\lambda$6355 expansion velocity of SN\,2024gsa by fitting the feature with a Gaussian function, deriving a velocity of $v_{\rm {Si~II}} = (16.7 \pm 1.1) \times 10^3$\,km\,s$^{-1}$, which is relatively high for SNe~Ic-BL at a similar phase\cite{Xiang2019}.

\noindent\textbf{Radio observations.} 
The radio follow-up observations of SN\,2024gsa were conducted with the Australia Telescope Compact Array (ATCA) on May 3, from 12:39:29.9 to 15:46:39.9, under project code CX568 (PI Tao An). The target had an on-source (exposure) time of 80\,min. The array was configured in its most extended 6\,km setup (6A). Data were simultaneously recorded at central frequencies of 5.5\,GHz and 9.0\,GHz, each with a bandwidth of 2\,GHz. The source position was determined using a flux density weighted mean of the peak values obtained at these two frequencies. The object is located at a right ascension (R.A.) of $12^h46^m01.682^s \pm 0.003^s$ and a declination (Dec.) of $-09^\circ43'08.13'' \pm 0.33''$. The radio observation positioning uncertainties arise from Gaussian fitting uncertainties and thermal noise. 
The overall accuracy of the source location is calculated using error propagation formulae and averaged from the measurement of the two frequencies. 

Standard data reduction was performed using the Common Astronomy Software Application (CASA). This process included flagging, calibration, and imaging. 
We created individual images for the 5.5\,GHz and 9.0\,GHz datasets to derive the spectral index.

The target was detected in ATCA C-band observations. The measured flux densities at various frequencies are presented in Extended Data Table \ref{table:flux_density} and Extended Data Figure \ref{fig:radio}. The source exhibits a rising spectrum with a spectral index of $+0.85 \pm 0.02$ between 5.5 and 9\,GHz, which is consistent with synchrotron self-absorption (SSA) radiation.

\subsection{Host-galaxy properties.\\}
The spectrum of J1246 is consistent with a redshifted Seyfert 1 AGN displaying typical emission lines, including H$\alpha$, H$\beta$, [O~III] $\lambda$5007, and [N~II] $\lambda$6583 (see Fig. \ref{fig:three_images} d). We fitted the emission lines in the observed spectrum of J1246 using a Gaussian model to determine the redshift. The fitting was performed with the SCIPY\cite{2020NatMe..17..261V} package. The redshift and its uncertainty were derived from the mean and standard deviation of the fitted Gaussian, resulting in $z = 0.401\pm0.003$. The redshifts of SN\,2024gsa and J1246 are consistent within the uncertainties. Therefore, SN\,2024gsa is likely associated with J1246 or one of its undetected satellite dwarf galaxies.
In either case, the host-galaxy extinction for SN\,2024gsa is expected to be minimal, while the Galactic extinction is $E(B-V)=0.032$\,mag\cite{2011ApJ...737..103S}. We adopt \( z = 0.401 \) for the analysis. 
Utilising the cosmological parameters from the Planck Collaboration model\cite{Planck2020}, we determine the corresponding luminosity distance to be \( 2245 \pm 10 \)\,Mpc, the angular diameter distance to be \( 1144 \pm 5 \)\,Mpc, and the distance modulus to be \( 41.76 \pm 0.01 \) mag. Therefore, the projected offset between SN\,2024gsa and the centre of J1246 is \(26.3 \pm 0.1\)\,kpc, significantly larger than the typical offset for SNe~Ic-BL\cite{Japelj2018} (Extended Data Figure \ref{fig:offset}).

To determine the metallicity of J1246, we utilised the O3N2 method based on flux ratios of specific emission lines in the optical spectrum. The spectral analysis was performed using the $\rm DASpec$ package (\url{https://github.com/PuDu-Astro/DASpec})\cite{Du2024}. 
The O3N2 parameter is defined as
\begin{equation}
\mathrm{O3N2} = \log_{10} \left( \frac{[\mathrm{O\,\textsc{iii}}]\,\lambda5007\,/\,\mathrm{H}\beta}{[\mathrm{N\,\textsc{ii}}]\,\lambda6584\,/\,\mathrm{H}\alpha} \right)\, .
\end{equation}
\noindent
Using the observed fluxes of these emission lines, we calculated the O3N2 value and derived the oxygen abundance following the calibration by\cite{Pettini2004},
\begin{equation}
12 + \log(\mathrm{O}/\mathrm{H}) = 8.73 - 0.32 \times \mathrm{O3N2}\, .
\end{equation}
Our measurements yielded a metallicity of \( 12~+~ \log(\mathrm{O}/\mathrm{H}) = 8.55~\pm~0.02 \)\, . 
Since the [O\,\textsc{iii}] and H\(\beta\) lines, as well as the [N\,\textsc{ii}] and H\(\alpha\) lines, are in close proximity in wavelength, the differential extinction between them is minimal. Therefore, we did not apply extinction correction to the emission-line fluxes.
We note that the SN is located in the outskirts of J1246. If it is the host galaxy of SN\,2024gsa, the metallicity at the SN's location is expected to be lower than the value we measured near the nucleus. Consequently, the metallicity at the location of SN\,2024gsa is lower than that of most SNe~Ic-BL\cite{Graham2013,Qin2024}.

The DESI Legacy Imaging Survey\cite{DESI2019} did not detect any objects at the position of SN\,2024gsa. The 5$\sigma$ limiting absolute magnitudes at the luminosity distance of SN\,2024gsa are $-$17.71 (\textit{g} band), $-$17.85 (\textit{r} band), $-$18.05 (\textit{i} band), and $-$18.47 (\textit{z} band). 
By converting the apparent magnitudes obtained from SDSS\cite{SDSS2015}, we derived that the absolute magnitudes of J1246 are $-21.68\pm0.03$ (\textit{g}), $-22.72\pm0.02$ (\textit{r}), $-23.17\pm0.02$ (\textit{i}), and $-23.56\pm0.05$ (\textit{z}). In the images taken with the Keck~I telescope, the galaxy exhibits a prominent central core accompanied by two distinct spiral arms (inset of Extended Data Figure \ref{fig:offset}). We classify it as a spiral galaxy with an active nucleus. An additional object near SN\,2024gsa, identified as Object 3 in the inset of Extended Data Figure \ref{fig:offset}, is a star in the Milky Way, as confirmed by Gaia distance measurements \cite{Gaia2021}.

We estimate the probability of chance coincidence ($P_{\text{ch}}$) between EP240414a/SN\,2024gsa and J1246. Adopting an apparent magnitude of $m_i = 18.2$ from the DESI Legacy Imaging Survey\cite{DESI2019} for the galaxy and an angular separation of $r = 4.75''$, we obtain $P_{\text{ch}} < 0.01$\cite{Bloom2002}. This result is consistent with the probability estimate presented by Ref. \cite{Srivastav2024arXiv}. This low probability suggests that the supernova is unlikely to be a chance projection.
Furthermore, the redshift of SN\,2024gsa ($z = 0.38 \pm 0.02$) is consistent with that of J1246 ($z = 0.401 \pm 0.003$), further reinforcing their physical association. 
Deep imaging from the Keck~I telescope reveals several other sources in the vicinity of SN\,2024gsa. However, the probabilities of chance coincidence for these sources are significantly higher than that of J1246. 
Given the low probability of chance coincidence and the redshift consistency, we conclude that SN\,2024gsa is most likely associated with J1246. However, the possibility of an association with an undetected satellite dwarf galaxy of J1246 cannot be ruled out.

\subsection{Spectral Analysis.\\}
\noindent\textbf{EP-WXT.} The integrated spectrum of the WXT prompt emission in the $T_{90}$ time interval (ranging from 19\,s to 174\,s) was fitted in \textit{XSPEC} by an absorbed power-law model \textit{tbabs*ztbabs*powerlaw}, where the first and second components are responsible for the Galactic absorption and the intrinsic absorption $N_{\rm int}$, and the third component is a power-law function, \(N(E) = K\times E^{\alpha}\), in the observer's frame. The column density of the Galactic absorption in the direction of EP240414a is fixed at \(3.35 \times 10^{20}\,\mathrm{cm}^{-2}\) \cite{Willingale2013} and the redshift is fixed at 0.401. A time-averaged intrinsic absorption of $N_{\rm int} = 7.4^{+4.1}_{-3.7} \times 10^{21}$\,$\rm cm^{-2}$ and a photon index $\alpha = -3.1_{-0.8}^{+0.7}$ are obtained with an acceptable statistic CSTAT/(d.o.f.) $\approx 6.66/13$ (Extended Data Figure \ref{fig:wxt_spec}a). The best-fit values of the photon index and intrinsic absorption and confidence contours are shown in Extended Data Figure \ref{fig:wxt_spec}b. The spectral energy distribution and the flux upper limits inferred from the  Konus Wind and {\it Swift}/BAT are shown in Extended Data Figure \ref{fig:wxt_spec}c. Furthermore, we consider an absorbed blackbody model \textit{tbabs*ztbabs*bbody}. The intrinsic absorption cannot be well constrained. An upper limit at the 90\% confidence level is given as $N_{\rm int} < 3.9 \times 10^{21}$\,$\rm cm^{-2}$ with this model. The temperature obtained is $0.35^{+0.04}_{-0.04}$\,keV. The statistic is not significantly improved in comparison with that fitted with the power-law model. The fitting results obtained and the corresponding fitting statistics are presented in Extended Data Table \ref{tab:spectral_analysis}. 

The soft spectrum fitted with the power-law model indicates that the energy peak $E_{\rm peak}$ is near or below 0.5\,keV, the lower limit of WXT's energy range. To place a quantitative constraint on the $E_{\rm peak}$, we fitted the spectrum with an absorbed broken power-law model \textit{tbabs*ztbabs*bknpower}. The first power-law index was fixed at -1, which is the typical value for GRBs. The second power-law index is fitted to be $-2.9_{-0.9}^{+0.6}$, which is consistent with the photon index derived from the fitting result with the single power-law model. The peak energy cannot be well constrained. An upper limit at the 90\% confidence level for the $E_{\rm peak}$ is obtained as 1.3\,keV. The fitted result gives a comparable statistic with an additional parameter.

Given the limited number of counts detected, we study the spectral evolution by decomposing the light curves into three smaller time slices with approximately equal net counts. We generated the source and background spectra, and fitted them with the absorbed power-law model. The intrinsic absorption is fixed at the same value as that derived from the power-law fit. The results in Extended Data Table \ref{tab:spectral_analysis} suggest that the photon index in the first time interval is softer than that in the later two time intervals. However, it is not statistically significant enough to claim a clear spectral evolution during the WXT observation.

\noindent\textbf{EP-FXT and Swift/XRT.} An X-ray source coincident with the optical position was detected in the first two FXT observations. The spectra (both FXT-A and FXT-B) can be fitted simultaneously by the same absorbed power-law model applied to the WXT spectra with an intrinsic absorption of $N_{\rm int}=2.4_{-2.0}^{+2.2}\times 10^{21}\,{\rm cm^{-2}}$ and a photon index of $-2.2^{+0.3}_{-0.4}$. In the seven XRT follow-up observations of the {\it Swift} XRT, a weak signal was detected only in the first epoch, with a significance of 5.2$\sigma$ as evaluated by the Li-Ma formula\cite{LiMa1983ApJ}. The absorbed flux is derived assuming a counts-to-flux ratio of $3.3\times 10^{-11}$\,$\rm erg\,cm^{-2}\,ct^{-1}$ with the HEASARC tool WebPIMMS\cite{Mukai1993}.

The upper flux limits at 90\% confidence level for the third FXT observation and the remaining XRT observations are inferred under the assumption of the same results from the FXT spectral fitting.

\subsection{Theoretical Modelling.\\}  
Based on the multiband observations, the evolution of EP240414a/SN\,2024gsa emission after the WXT detection can be characterised by three distinct physical processes. Detailed analysis and modelling for each phase are presented below.

\noindent\textbf{Afterglow of a relativistic jet interacting with surrounding medium (Phase 1).}
The simultaneous decay of X-ray and optical fluxes, from approximately $10^4$\,s to $10^5$\,s, can be attributed to the afterglow of a successful jet. Using the classical GRB afterglow model \cite{sari1998, gao2013} and fitting the multiband light curves with $F_{\nu} \propto t^{-\alpha}\nu^{-\beta}$, we find $\alpha \approx 0.3$ and $\beta \approx 0.9$.  Assuming the Lorentz-factor distribution for shock-accelerated electrons follows \(dN(\gamma_e) \propto \gamma_e^{-p} d\gamma_e\) \cite{sari1998, meszaros1998}, the spectrum suggests \(p = 2.8\) which is within the slow cooling regime. The temporal index indicates the jet is in the self-deceleration phase with an energy injection. For an injection rate $L_{\rm inj} \propto t^{-q}$ (Ref.\cite{zhang&meszaros2001}), $q = 0.28$ is obtained. The radio observation, taken 19 days after $T_0$, can also be explained by the jet afterglow. While the narrow radio spectrum is not sufficient to determine the exact spectral index, the flux clearly increases with frequency. Thus, we have ${\rm max}(\nu_m, \nu_a) > 9.0\,{\rm GHz}$, where $\nu_a$ and $\nu_m$ are the self-absorption frequency and the synchrotron frequency related to the accelerated electrons at the low-energy end, respectively. By this time, the energy injection is assumed to have stopped.

We have tried to constrain the parameters for the jet-medium interaction, including the isotropic kinetic energy ($E_{\rm k,iso}$), the CSM density ($n_0$), and the fraction of energy converted to the random motion of electrons ($\epsilon_e$) and magnetic field ($\epsilon_B$). Analytical expressions for the characteristic frequencies and fluxes are used\cite{gao2013}. 
It is challenging for a model with a homogeneous medium (constant $n_0$) and constant energy conversion efficiencies ($\epsilon_e$ and $\epsilon_B$) to fully explain the multi-band afterglows observed in X-ray, optical, and radio.
We assume an evolving CSM and treat the early (X-ray/optical) and late (radio) afterglows separately. Despite limited data, it is possible to find parameters that explain the observations. The best-fit parameters are $n_{0,1} = 1.0\,{\rm cm^{-3}}$, $\epsilon_{e,1} = 0.01$, $\epsilon_{B,1} = 0.0001$ for the early phase, and $n_{0,2} = 0.3\,{\rm cm^{-3}}$, $\epsilon_{e,2} = 0.5$, $\epsilon_{B,2} = 0.05$ for the late phase, with $E_{\rm k,iso} = 1.0\times 10^{51}\,{\rm erg}$. The results of the fit are presented in Extended Data Figure \ref{fig:afterglow_fit}. The jet's deceleration time ($t_{\rm dec}$)\cite{gao2013} must be shorter than the first afterglow observation at $t_s = 1.3 \times 10^4\,{\rm s}$, setting a rough lower limit for the jet's initial bulk Lorentz factor ($\Gamma_0$) at $\sim 13$. The radio luminosity ($\nu L_{\nu}$) of SN\,2024gsa is compared with that of other transient events in Extended Data Figure \ref{fig:radio}. The $\nu L_{\nu}$ of SN\,2024gsa is significantly brighter than that of SNe and fast blue optical transients (FBOTs) at similar phases, and is comparable to that of luminous GRBs, such as GRB\,030329. This further suggests that the radio emission originates from the afterglow of a jet, rather than from the radio emission of a supernova.

\noindent\textbf{Shock-cooling model for the optical peak during Phase 2.}
The optical light curve of SN\,2024gsa during Phase 2 is featured by rapid rise and fall times, along with high luminosities. These fast variations and bright peak suggest that the peak cannot be powered by the radioactive decay of \(^{56}\)Ni, which is the primary energy source for many core-collapse SNe, nor by cooling emission after shock breakout from the surface of a supergiant, as the latter typically occurs on much shorter timescales\cite{Li2024Natur}. One potential mechanism for generating such a rapid and luminous light curve is to deposit energy into dense, extended material at large radii\cite{Ofek2010, Chevalier2011}. If this dense material possesses a sufficiently high optical depth near its outer edge (\(R_{\text{ext}}\)), radiation will first break out when the shock reaches \(R_{\text{ext}}\). In this scenario, the optical peak is primarily driven by shock breakout and cooling emission from the expanding hot material, rather than by ongoing shock interaction\cite{Margalit2022, Khatami2024}.

We model the early light curve of SN\,2024gsa as cooling emission from shock-heated extended material surrounding the progenitor.
We apply the shock cooling model from a dense CSM framework developed by \cite{Margalit2022} to fit the Phase 2 optical light curve. This model improves upon earlier shock-cooling emission frameworks \cite{Piro2015, Piro2021} by explicitly solving the radiative diffusion equation, which accounts for energy transport and losses within the extended material. The rise time of the emission is mainly determined by the mass of the extended material (\(M_{\rm ext}\)), while the peak luminosity is influenced by its radius (\(R_{\rm ext}\)). Our modeling yields an envelope radius of \(R_{\rm ext} = 2.41^{+0.51}_{-1.03} \times 10^{14}\)\,cm and a mass of \(M_{\rm ext} = 0.33^{+0.06}_{-0.03}\,M_{\odot}\). These results support a scenario in which a compact SN~Ic-BL progenitor experienced eruptive mass loss shortly before the explosion, with the light curve initially dominated by shock breakout and post-shock cooling of the recently ejected material. For comparison, SN\,2006aj/GRB\,060218 was modeled as a shock breakout from a low-mass (\(0.01\,M_{\odot}\)) extended envelope with a large radius (\(100\,R_{\odot}\)), powered by a low-luminosity GRB\cite{Nakar2015}. Other SNe~Ic-BL, such as iPTF14gqr and iPTF16asu, also showed early signs of interaction driven by post-shock cooling, but with lower masses and smaller radii in their extended material, such as \(M_{\text{ext}} \approx 8 \times 10^{-3}\,M_{\odot}\) and \(R_{\text{ext}} \approx 3 \times 10^{13}\)\,cm for iPTF14gqr\cite{De2018}, and \(M_{\text{ext}} \approx 0.45\,M_{\odot}\) and \(R_{\text{ext}} \approx 1.7 \times 10^{12}\)\,cm for iPTF16asu\cite{Whitesides2017}.

\noindent\textbf{Nickel-powered model for the peak during Phase 3.} 
Given the similarities in the optical spectrum of SN\,2024gsa with those of typical SNe~Ic-BL, we propose that the peak during Phase 3 is powered by the radioactive decay of \(^{56}\)Ni. To model this, we applied a simple Arnett model \cite{Arnett1982} to fit the observed optical light curve, where the key parameters are the nickel mass, ejecta mass, and ejecta velocity. Using the Markov Chain Monte Carlo (MCMC) method implemented in the Python \textit{emcee} package \cite{Foreman2013}, we fit the multiband light curve during Phase 3. The fitting results are shown in Figure \ref{fig:lightcurve}. Our best-fit model yields an ejecta mass of \(M_{\rm{ej}}=2.38^{+0.45}_{-0.35}\,M_{\odot}\), a nickel mass of \(M_{\rm{Ni}}=0.74^{+0.05}_{-0.04}\,M_{\odot}\), and an ejecta velocity of \(v_{\rm{ej}}=1.5^{+0.21}_{-0.19}\times 10^{4}\)\,km\,s\(^{-1}\).
 
These derived parameters suggest that SN\,2024gsa is consistent with the characteristics of typical SNe Ic-BL, particularly in terms of ejecta mass and velocity, which fall within the typical ranges for this SN type \cite{Cano2017}. Moreover, these values are also typical of GRB-associated supernovae, such as SN 1998bw and SN 2006aj. Specifically, the inferred parameters for SN 1998bw include an ejecta mass of $M_{\text{ej}}\approx (6-10) M_{\odot}$, a nickel mass of $M_{\text{Ni}} \approx (0.3-0.6) M_{\odot}$, and an expansion velocity of $v_{\text{ej}} \approx 18, 000 \text{km s}^{-1}$ \cite{Galama1998Natur}, while for SN 2006aj, the corresponding values are $M_{\text{ej}} =(2.0\pm 0,5) M_{\odot}$, $M_{\text{Ni}}\approx (0.2\pm 0.1)  M_{\odot}$, and $v_{\text{ej}} \approx 20, 000 \text{km s}^{-1}$ \cite{Pian2006Nature}
The relatively high nickel mass inferred from our model indicates a significant amount of synthesised radioactive material, which is crucial in explaining the brightness of the Phase 3 light-curve peak.

\subsection{Event rate density.\\}
The unique properties of EP240414a suggest a different physical origin from SN shock breakout and low-luminosity GRBs. With one detection, we estimate the local event rate density $\rho_{0,\rm EFXT}$ following the method in Ref\cite{Sun2015}. The detected number of the event $N_{\rm EFXT}$ can also be given through
\begin{equation}
    N_{\rm EFXT} = \frac{\eta\,\Omega_{\rm WXT}\,T_{\rm OT}}{4\pi}\rho_{0,\rm EFXT}\,V_{\rm max} = 1\,.
\end{equation}
The field of view of EP-WXT, $\Omega_{\rm WXT}$, is 3600 square degrees. The operation time of EP-WXT $T_{\rm OT}$ by this work is $\sim 8$ months with a duty cycle $\eta =50\%$. The sensitivity of EP-WXT with a typical exposure time of 200\,s reaches $\sim 1 \times 10^{-10}$\,$\rm erg\,cm^{-2}\,s^{-1}$ in the 0.5--4\,keV band. With a peak luminosity of $1.3 \times 10^{48}$\,$\rm erg\,s^{-1}$, the maximum redshift that the source can be detected at a signal-to-noise ratio of 7 is $z_{\rm max} = 0.7$. 

The $V_{\rm max}$ corresponds to the effective maximum volume within which the event can be detected. The function $f(z)$ characterizes the evolution of the event rate density with redshift and is modeled based on the star-formation history\cite{Yuksel2008ApJ} for EP240414a.
\begin{equation}
V_{\rm max} = \int_{0}^{z_{\rm max}} \frac{\Omega_{\rm WXT}}{4\pi}\,\frac{f(z)}{(1+z)} \frac{dV(z)}{dz} dz.
\label{eq:V'max}
\end{equation}
The comoving volume is given by
\begin{equation}
\frac{dV(z)}{dz}=\frac{c}{{\rm H}_0}\frac{4\pi D_L(z)^2}{(1+z)^2[\Omega_M(1+z)^3+\Omega_{\Lambda}]^{1/2}}\, ,
\end{equation} 
where $D_L(z)$ is the luminosity distance at the corresponding redshift $z$. We assume a flat $\Lambda$CDM cosmology with $H_0 = 67.4$~km~s$^{-1}$~Mpc$^{-1}$ and $\Omega_M = 0.315$\cite{Planck2020}.

The above calculation resulted in a local event rate density of
\begin{equation}
\rho_{0,\rm EFXT}\approx 0.3^{+0.7}_{-0.2} \,\rm Gpc^{-3} \, yr^{-1}\, .
\end{equation} 
The error represents 1-$\sigma$ uncertainty, which is calculated from small-sample statistics\cite{Gehrels1986ApJ}. It is crucial to acknowledge that this rate is the lower limit of the intrinsic value, given that EP has detected more than thirty fast X-ray transients without gamma-ray counterparts (Wu et al. in prep.). The physical origins of these transients remain uncertain owing to the lack of timely follow-up observations and reliable multiband confirmation. Under the assumption that these events have an origin similar to that of EP240414a, we obtain an upper limit on the event rate density of $< 10$ $\rm Gpc^{-3} \, yr^{-1}$.

\clearpage

\section*{Data Availability}
The light curves and spectra of EP-WXT and EP-FXT and the spectroscopic data are available at https://github.com/huisungh/EP240414a.git. The light curves of Swift-BAT GRBs are public and
can be found at https://www.swift.ac.uk/burst\_analyser.

\section*{Code Availability}
Upon reasonable requests, the code (mostly in Python) used to produce the results and figures will be provided.

\bigskip
\bigskip
\bigskip


\clearpage

\begin{addendum}

\item[Acknowledgments] This work is based on data obtained with the Einstein Probe (EP), a space mission supported by the Strategic Priority Program on Space Science of the Chinese Academy of Sciences, in collaboration with ESA, MPE, and CNES (grant XDA15310000); the Strategic Priority Program on Space Science of the Chinese Academy of Sciences (grant No. E02212A02S) and the Strategic Priority Research Program of the Chinese Academy of Sciences (Grant No. XDB0550200). 
We acknowledge the support by the National Natural Science Foundation of China (NSFC grants 12288102, 12373040, 12021003, 12103065, 12333004, 12203071, 12033003, 12233002, and 12303047).
This work is also supported by the National Key R\&D Program of China (grant 2022YFF0711500).
W.X.L, S.J.X., H.Z., and W.J.G. acknowledge the supports from the Strategic Priority Research Program of the Chinese Academy of Sciences (grant Nos. XDB0550100 and XDB0550000), National Key R\&D Program of China (grant Nos. 2023YFA1607804, 2022YFA1602902, and 2023YFA1608100), and National Natural Science Foundation of China (NSFC; grant Nos. 12120101003, 12373010, and 12233008)
X.W's group at Tsinghua University is supported by NSFC (grants 12288102 and 12033003), and the Tencent Xplorer Prize. A.V.F.'s group at UC Berkeley is grateful for financial assistance from the Christopher R. Redlich Fund,
Gary and Cynthia Bengier, Clark and Sharon Winslow, Alan Eustace (W.Z. is a Bengier-Winslow-Eustace Specialist in Astronomy), William Draper, Timothy and Melissa Draper, Briggs and Kathleen Wood, Sanford Robertson (T.G.B. is a Draper-Wood-Robertson Specialist in Astronomy), and many other donors. S.A. has received support from the Carlsberg Foundation (CF18-0183, PI: I. Tamborra). This work is supported by the ANID FONDECYT project No. 3220029. Z.G. is funded by ANID, Millennium Science Initiative, AIM23-001. 
Partly based on observations made with the Nordic Optical Telescope, owned in collaboration by the University of Turku and Aarhus University, and operated jointly by Aarhus University, the University of Turku and the University of Oslo, representing Denmark, Finland and Norway, the University of Iceland and Stockholm University at the Observatorio del Roque de los Muchachos, La Palma, Spain, of the Instituto de Astrofisica de Canarias.
A.J.C.T. acknowledges support from the Spanish Ministry projects PID2020-118491GB-I00 and PID2023-151905OB-I00 and Junta de Andaluc\'ia grant P20\_010168 and from the Severo Ochoa grant CEX2021-001131-S funded by MCIN/AEI/ 10.13039/501100011033.
We acknowledge the support of the staffs of the 10.4~m Gran Telescopio Canarias (GTC) and Keck~I 10~m telescope. This work makes use of the Las Cumbres Observatory global network of robotic telescopes.  The LCO group is supported by NSF grants AST-1911225 and AST-1911151. S.B. and N. Elias-Rosa acknowledges support from the PRIN-INAF 2022,
`Shedding light on the nature of gap transients: from the observations to the models'. We gratefully acknowledge the China National Astronomical Data Center (NADC), the Astronomical Data Center of the Chinese Academy of Sciences, and the Chinese Virtual Observatory (China-VO) for providing data resources and technical support. The work of D.S.S., A.V.R., D.D.F, was supported by the basic funding programme of the Ioffe Institute no. FFUG-2024-0002.

\item[Author Contributions] W.Y. has been leading the Einstein Probe project as Principal Investigator since the mission proposal stage.  H.G., H.S., B.Z., W.X.L., X.F. Wang, and Y.L. initiated the study. H.G., X.F. Wang, B.Z., X.F. Wu, H.S., W.X.L., and L.D.L. coordinated the scientific investigations of the event and led the discussions. H.S., Y.L., T.Y.L. processed and analysed the WXT data. Q.-Y.W. and H.S. processed and analysed the FXT data. Q.-Y.W. processed and analysed the XRT data. W.X.L., X.F. Wang, and D.X. led the optical and near-infrared data taking and data analysis. A.V.F., W.Z. Y.Y., T.G.B., K.S.T., N. Elias-Rosa obtained and reduced the optical/NIR spectroscopy and photometry. T.A. and Y.Q.L. helped with the radio data taking and analysis. S.B. helped with the reduction of GTC photometry and spectroscopy. D.M., S.A.E., and A.H. contributed to the optical data taking with AZT-22 telescope and LCO 1-m telescopes, respectively. L.D.L., H.G., B.Z., S.A. led the theoretical investigation of the event. W.X.L, H.S., B.B.Z., X.F. Wang, D.X. contributed to the theoretical investigation of the event. C.Y.W., B.B.Z., B.Z. contributed to comparing this event with GRBs. H.S. contributed to the event rate density. J.D. performed the GRB search in Swift/BAT data and the upper limit. D.S.S., A.V.R. and D.D.F. performed GRB search in the Konus-Wind data and the upper limit. T.-Y.L., X.P., Y.-F.L., J.Y., and C.-Y.D. are the transient advocates on April 14 2024 and contributed to the discovery and preliminary analysis of the event.
Z.-X.L., C.Z., S.-N.Z., X.-J.S., S.-L.S., X.-F.Z., Y.-H.Z., Z.-M.C. F.-S.C. and W.Y. contributed to the development of the WXT instrument. C.Z., Z.-X.L., H.-Q.C., D.-H.Z. and Y.L. contributed to the calibration of WXT data.
Y.L., H.-Q.C., C.J., W.-D.Z., D.-Y.L., J.-W.H., H.-Y.L., H.S., H.-W.P. and M.-J.L. contributed to the development of WXT data analysis software.
Y.C., S.-M.J., W.-W.C., C.-K.L., D.-W.H., J.W., W.L., Y.-J.Y., Y.-S.W., H.-S.Z., J.G., J.Z., X.-F.Z., J.-J.X., J.M., L.-D.L., H.W., X.-T.Y., T.-X.C., J.H., Z.-J.Z., Z.-L.Z., M.-S.L., Y.-X.Z., D.-J.H., L.-M.S., F.-J.L., C.-Z.L., Q.-J.T. and H.-L.C. contributed to the development of the FXT instrument. 
S.-M.J., H.-S.Z., C.-K.L., J.Z., and J.G. contributed to the development of FXT data analysis software. 
W.X.L., H.S., L.D.L., H.G., X.F. Wang, B.Z., and S.A. drafted the manuscript with the help from all authors. A.V.F. assisted with editing the manuscript. 

\item[Competing Interests] The authors declare no competing interests.

\end{addendum}

\clearpage

\setcounter{figure}{0}
\setcounter{table}{0}
\captionsetup[figure]{labelfont={bf}, labelformat={default}, labelsep=period, name={Extended Data Fig.}}
\captionsetup[table]{labelfont={bf}, labelformat={default}, labelsep=period, name={Extended Data Table}}

\begin{figure}
\center
\begin{overpic}[width=0.45\textwidth]{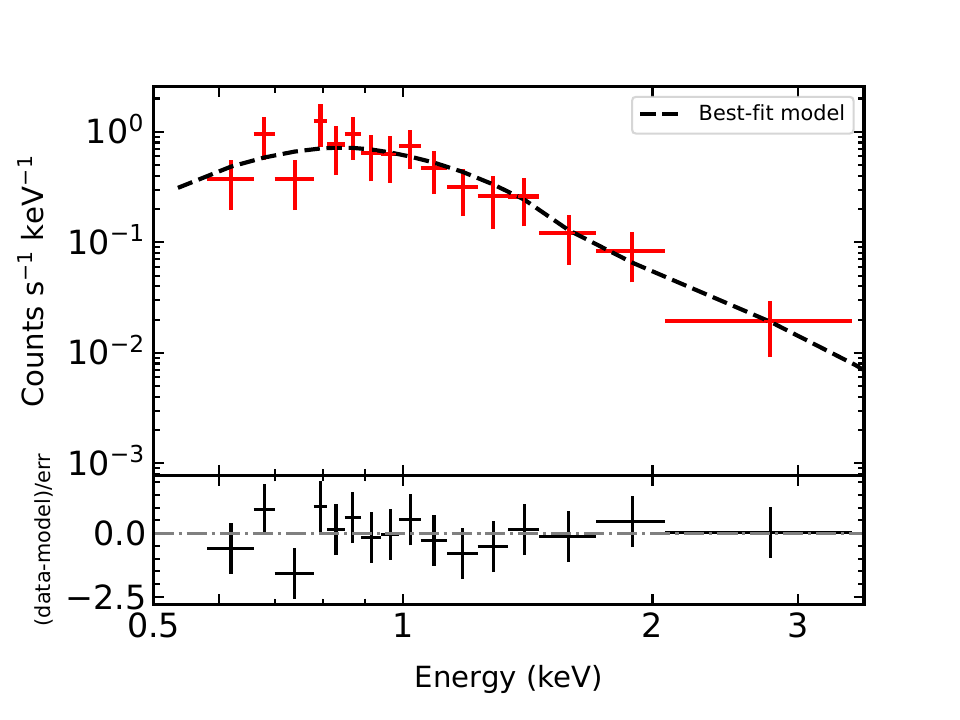}\put(1, 66){\bf a}\end{overpic} 
\begin{overpic}[width=0.45\textwidth]{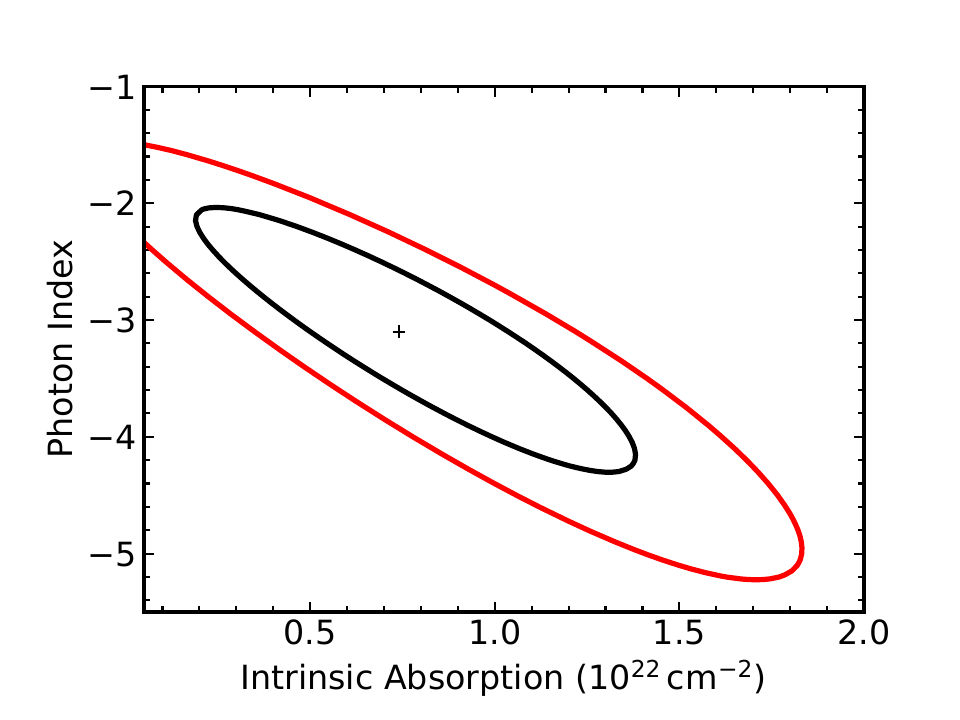}\put(1, 66){\bf b}\end{overpic} \\
\begin{overpic}[width=0.5\textwidth]{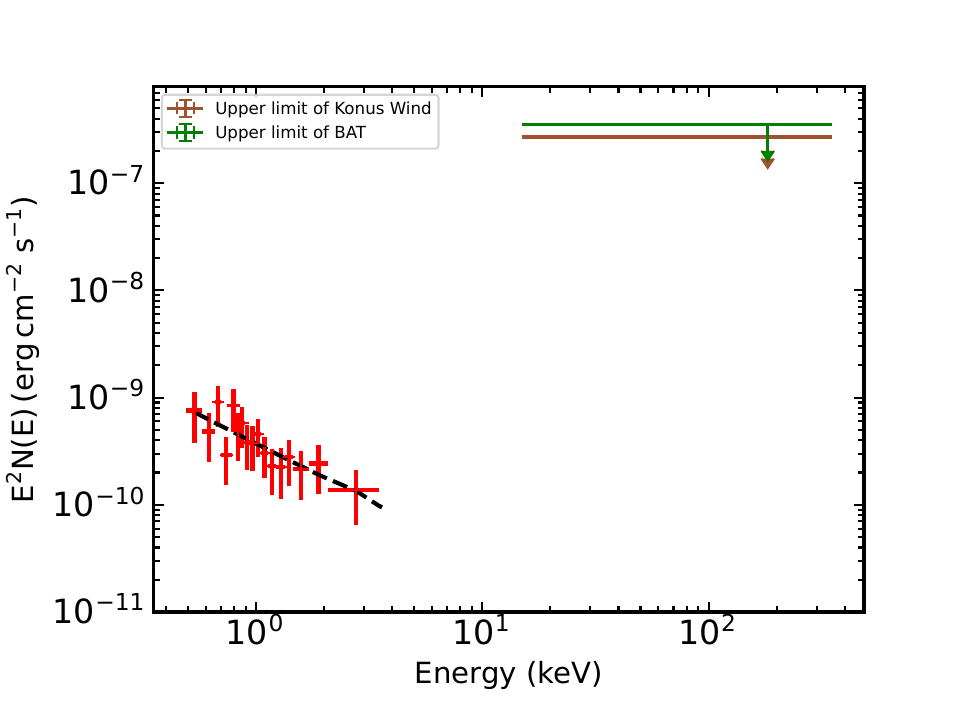}\put(1, 65){\bf c}\end{overpic} \\
\caption{\textbf{The WXT spectrum in the time interval of $T_{90}$.} \textbf{a}, The WXT observed spectrum and the predicted best-fit absorbed power-law model. Data are presented as the count rate spectrum with 1$\sigma$ uncertainties. \textbf{b}, Best-fit values of photon index and intrinsic absorption $N_{\rm int}$ and the 1$\sigma$, 2$\sigma$ confidence contours. \textbf{c}, The spectral energy distribution. Data are presented as the energy flux density with 1$\sigma$ uncertainties. The upper limits displayed in brown and green represent those of Konus Wind and {\it Swift}/BAT, respectively.}
\label{fig:wxt_spec}
\end{figure}

\clearpage

\begin{figure}
\center
\includegraphics[width=1\textwidth]{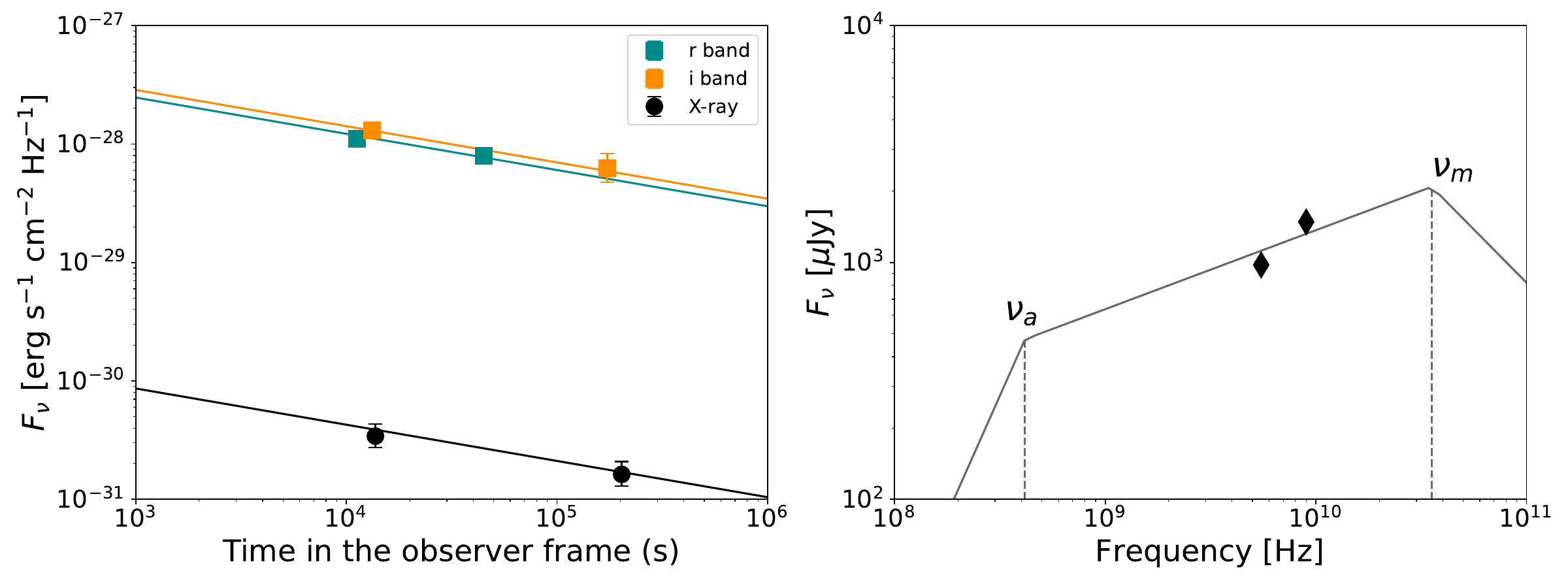}
\caption{\textbf{Fitting of the theoretical afterglow model to the X-ray, optical, and radio data.} 
Left panel: The early X-ray (1 keV) and optical (r and i band) afterglow light curves and the modeling with the classical GRB afterglow model \cite{sari1998, gao2013}. Data are presented as the measured flux density with 1$\sigma$ uncertainties. Right panel: The synchrotron spectrum in the radio band at $T_0+$19 days. The $\nu_a$ and $\nu_m$ are the self-absorption frequency and the synchrotron frequency related to the accelerated electrons at the low-energy end, respectively.}
\label{fig:afterglow_fit}
\end{figure}

\clearpage

\begin{figure}
\centering
\begin{tabular}{c}
\includegraphics[width=0.75\textwidth]{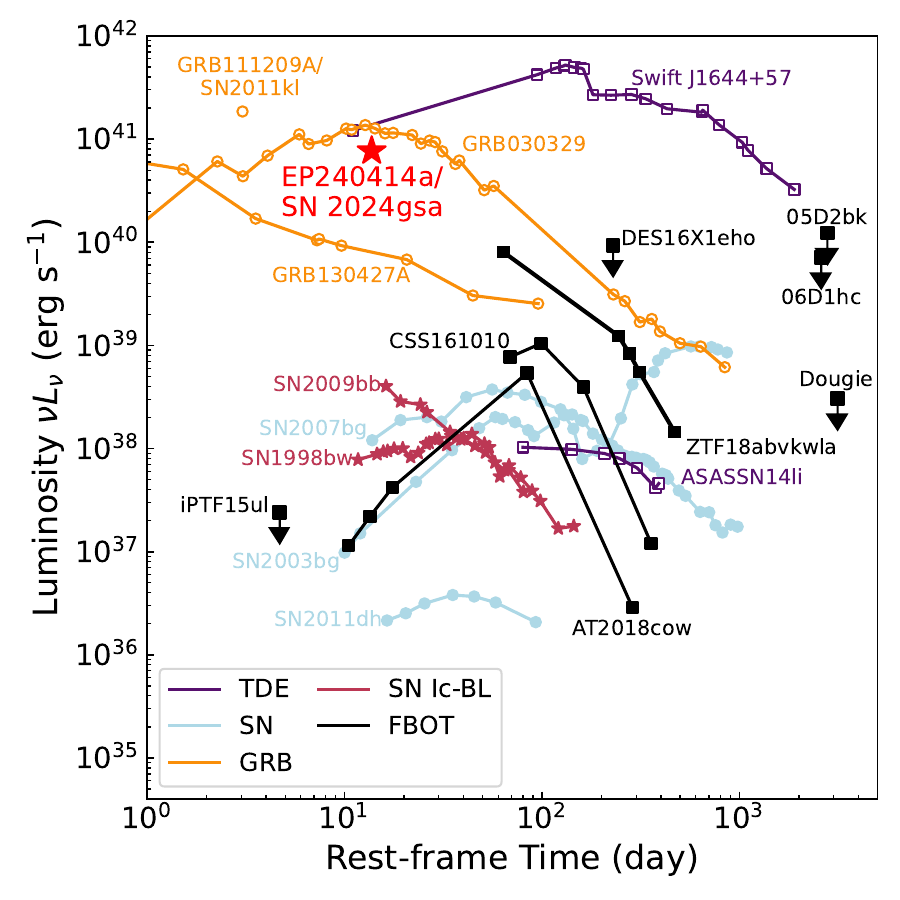}
\end{tabular}
\caption{\noindent\textbf{Radio luminosity of EP240414a/SN\,2024gsa.} The 9\,GHz radio luminosity of EP240414a/SN\,2024gsa is compared to low-frequency (1--10\,GHz) light curves of different classes of energetic explosions: tidal disruption events \cite{Berger2012,Alexander2016}, SNe \cite{Soderberg2005,Salas2013}, relativistic Ic-BL SNe\cite{Kulkarni1998,Soderberg2010}, long-duration GRBs \cite{Berger2003,Soderberg2006,Perley2014}, and fast blue optical transients \cite{Margutti2019,Coppejans2020,Ho2020}. }
\label{fig:radio}
\end{figure}

\clearpage

\begin{figure}
\centering
\includegraphics[width=0.80\textwidth]{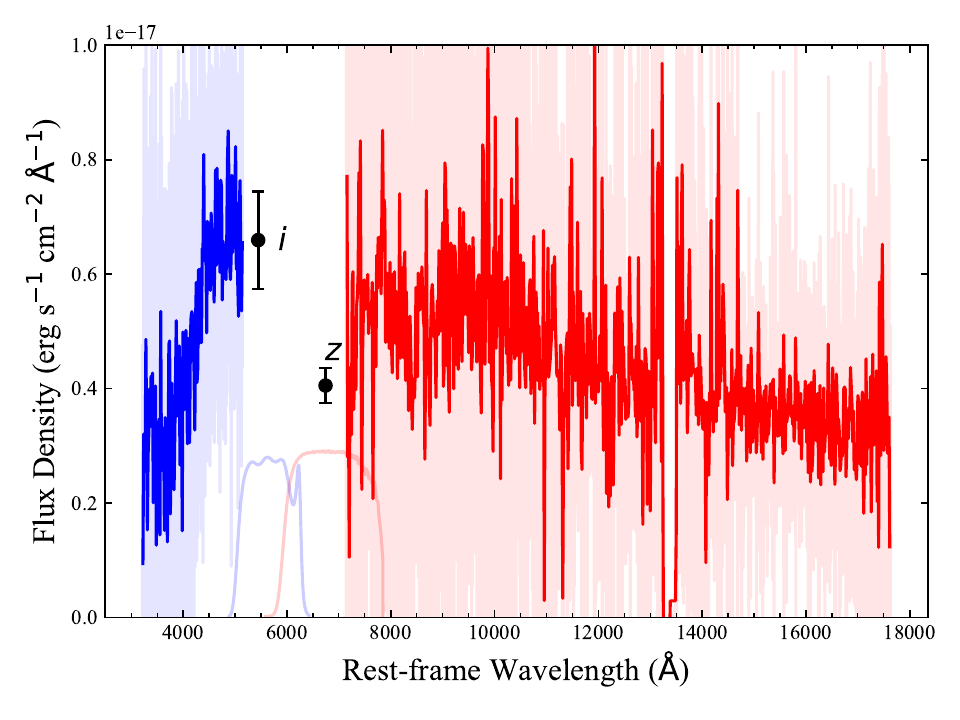}
\caption{\noindent\textbf{Redshift-corrected optical and NIR spectra of SN\,2024gsa.} The optical spectrum was obtained on April 19.9 using the GTC/OSIRIS+ instrument (light blue) and the NIR spectrum was obtained on April 19.3 using the Keck/NIRES instrument (light red). Rebinned versions of both spectra, generated using a bin width of 20 \AA, are also overplotted in blue and red, respectively. Two photometric data points taken at similar phases in the $i$ and $z$ bands, converted to flux density, are also plotted along with their transmission curves. The uncertainties associated with the two data points are shown at the 1$\sigma$ confidence level. The effective wavelengths have been redshift-corrected, while the flux density remains uncorrected for redshift.
}
\label{fig:GTC}
\end{figure}

\clearpage

\begin{figure}
\centering
\begin{tabular}{c}
\includegraphics[width=0.8\textwidth]{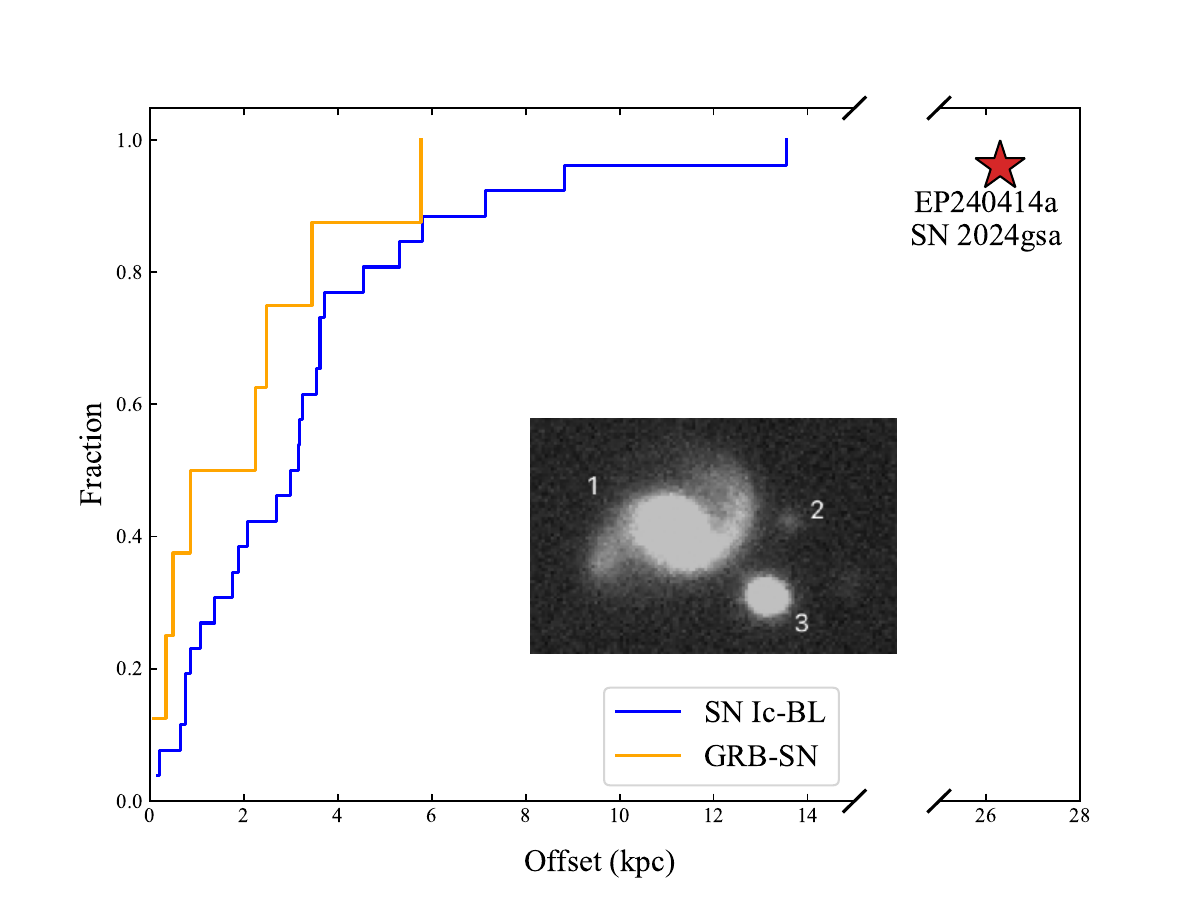}
\end{tabular}
\caption{\noindent\textbf{Projected offsets of SNe~Ic-BL and GRB-SNe.} The cumulative distributions of the projected offsets from the host-galaxy centres for a sample of SNe~Ic-BL (blue) and GRB-SNe (yellow) are shown as solid lines\cite{Japelj2018}, with SN\,2024gsa marked by a red star. An $I$-band image of the host galaxy, obtained with the Keck-I telescope, is overlaid as an inset;
Object 1 corresponds to J1246, the faint Object 2 to SN\,2024gsa, and Object 3 to a foreground point source.}
\label{fig:offset}
\end{figure}

\clearpage

\begin{table*}
\centering
\scriptsize
\begin{threeparttable}
\caption{Spectral results of X-ray observations of EP240414a. Errors represent the 1$\sigma$ uncertainties. The upper limits are at the 90\% confidence level.}\label{tab:spectral_analysis}
\begin{tabular}{llcccccccc}
\toprule
Instrument & Time Interval &  Model\tnote{*} & $\alpha$ & $\beta$ & $E_{\rm peak}$ & $ T_{\rm BB}$ & $ N_{\rm int}$ &   $F_{\rm abs}$  & CSTAT/(d.o.f.) \\
& &  & & & (keV) & (keV) & ($ 10^{21} \rm{cm^{-2}}$)  & ($\rm erg\,cm^{-2}\,s^{-1}$) &  \\
\hline
 
\multirow{5}{*}{EP-WXT} & [19\,s, 174\,s] &  PL  &  $-3.1_{-0.8}^{+0.7}$ & -- & -- & -- & $7.4^{+4.1}_{-3.7}$ & $6.5^{+1.3}_{-1.0} \times 10^{-10}$ & $6.66/13$ \\
& [19\,s, 174\,s] &  BB  &  -- & -- & -- & $0.35^{+0.04}_{-0.04}$ & $<3.9$ & $6.2^{+0.9}_{-0.9} \times 10^{-10}$ & $8.60/13$ \\
& [19\,s, 174\,s] &  BPL  & $-1$ (Fixed) & $-2.9_{-0.9}^{+0.6}$ & $< 1.3$ & --  & $4.0^{+6.2}_{-3.4}$ & $6.6^{+1.2}_{-1.1} \times 10^{-10}$ & $6.21/12$ \\
& [19\,s, 49\,s] &  PL  &  $-3.9^{+0.9}_{-0.9}$ &  -- & -- &-- & $7.4$ (Fixed) & $9.2^{+3.5}_{-2.2} \times 10^{-10}$ & $9.16/10$ \\
& [49\,s, 73\,s] &  PL  &  $-2.7^{+0.5}_{-0.6}$ &  -- & -- &-- & $7.4$ (Fixed) & $1.5^{+0.5}_{-0.4} \times 10^{-9}$ & $15.36/11$ \\
& [73\,s, 174\,s] &  PL  &  $-3.0^{+0.7}_{-0.7}$ &  -- & -- &-- & $7.4$ (Fixed) & $3.5^{+1.4}_{-0.9} \times 10^{-10}$ & $14.93/11$ \\
\hline
\multirow{3}{*}{EP-FXT} & [7.5\,ks, 20.1\,ks] & \multirow{3}{*}{PL} &  \multirow{3}{*}{$-2.2^{+0.3}_{-0.4}$} &  -- & -- &-- & \multirow{3}{*}{$2.4^{+2.2}_{-2.0}$} & $1.8^{+0.4}_{-0.3} \times 10^{-13}$ & \multirow{3}{*}{$115.66/97$}\\
&[2.27\,d, 2.44\,d] &  &   & -- & -- &--  &  & $8.3^{+1.9}_{-1.5} \times 10^{-14}$ & \\
&[45.14\,d, 45.31\,d] &    &    &  -- & -- & -- &  & $< 2.8 \times 10^{-14}$ & -- \\
\hline
\multirow{7}{*}{{\it Swift}/XRT} & [4.22\,d, 4.30\,d] & \multirow{7}{*}{PL}   &   &  -- & -- &-- & & $1.4^{+0.6}_{-0.6} \times 10^{-13}$ & -- \\
& [10.16\,d, 10.42\,d] &    &    &  -- & -- &-- &   & $< 7.7 \times 10^{-14}$ & -- \\
& [12.65\,d, 13.59\,d] &    &   &  -- & -- &-- & & $ <9.7 \times 10^{-14}$ & -- \\
& [21.00\,d, 21.29\,d] &    &  $-2.2$ &  -- & -- &-- & $2.4$  & $ <7.3 \times 10^{-14}$ & -- \\
& [30.61\,d, 30.67\,d] &    & (Fixed)  &  -- & -- &-- & (Fixed) & $ <1.5 \times 10^{-13}$ & -- \\
& [48.63\,d, 49.51\,d] &    &   &  -- & -- &-- &  & $ <9.9 \times 10^{-14}$ & -- \\
& [50.60\,d, 51.41\,d] &    &   &  -- & -- &-- &  & $ <4.0 \times 10^{-14}$ & -- \\
\bottomrule
\end{tabular}

\begin{tablenotes}
\footnotesize
\item[*] PL, BB, and BPL represent the power-law, blackbody, and broken power-law models, respectively.

\end{tablenotes}
\end{threeparttable}

\end{table*}

\clearpage

\begin{table*}
\centering
\small
\caption{Log of X-ray follow-up observations by EP-FXT and {\it Swift}/XRT.}\label{tab:FXT_XRT_obsinfo}
\begin{tabular}{lcccc}
\toprule
ObsID & Start time & End time & Exposure  \\
 & (UTC) &(UTC)& (s) \\
\hline
FXT & & & \\
\hline
08500000064 & 2024-04-14T11:54:10 & 2024-04-14T15:24:50 & 7218 \\
08500000068 & 2024-04-16T16:19:57 & 2024-04-16T20:24:01 & 9246 \\
08500000104 & 2024-05-29T13:13:02 & 2024-05-29T17:15:01 & 8934 \\
\hline
XRT & & & \\ 
\hline
00016609001 & 2024-04-18T15:02:27 & 2024-04-18T16:54:25  & 1865 \\
00016609002 & 2024-04-24T13:35:51 & 2024-04-24T19:58:53  & 1344 \\
00016609003 & 2024-04-27T01:28:59 & 2024-04-27T23:51:52  & 1702 \\
00016609004 & 2024-05-05T09:54:49 & 2024-05-05T16:40:53  & 4434 \\
00016609006 & 2024-05-15T00:34:58 & 2024-05-15T02:05:52  & 1877 \\
00016609007 & 2024-06-02T00:56:15 & 2024-06-02T22:02:53  & 5541 \\
00016609008 & 2024-06-04T00:17:44 & 2024-06-04T19:45:54  & 2822 \\
\bottomrule
\end{tabular}
\end{table*}

\clearpage
\begin{longtable}{cccc}
\caption{Optical and NIR photometry of SN\,2024gsa. Errors represent the 1$\sigma$ uncertainties. The \textit{griz}-band photometry is given in the AB magnitude system, while all other bands are in the Vega magnitude system.} \label{table:opt_ir} \\ 
\toprule
Time (day) & Magnitude  & Filter & Telescope \\ 
\midrule
\endfirsthead
\toprule
Time (day) & Mag  & Filter & Telescope  \\ 
\midrule
\endhead
\midrule
\multicolumn{4}{c}{{Continued on next page}} \\ 
\midrule
\endfoot
\bottomrule
\endlastfoot
0.23  & $>$ 19.3 & clear & BOOTES-4/MET\\
0.521	&	21.88 $\pm$ 0.07 	&	$r$	&	NOT	 \\
1.196	&	$>$ 20.50        	&	$R$	&	XLT	 \\
1.501	&	22.77 $\pm$ 0.21 	&	$g$	&	NOT	 \\
1.519	&	22.32 $\pm$ 0.19 	&	$i$	&	NOT	 \\
1.534	&	22.40 $\pm$ 0.15 	&	$r$	&	NOT	 \\
3.632	&	20.62 $\pm$ 0.10 	&	$r$	&	NOT	 \\
3.645	&	20.30 $\pm$ 0.25 	&	$z$	&	NOT	 \\
3.656	&	21.26 $\pm$ 0.24 	&	$B$	&	LCO	 \\
3.662	&	20.98 $\pm$ 0.23 	&	$V$	&	LCO	 \\
3.665	&	20.96 $\pm$ 0.16 	&	$g$	&	LCO	 \\
3.671	&	20.72 $\pm$ 0.19 	&	$r$	&	LCO	 \\
3.674	&	20.65 $\pm$ 0.14 	&	$i$	&	LCO	 \\
4.4	&	20.80 $\pm$ 0.30 	&	$r$	&	LJT	 \\
4.42	&	20.37 $\pm$ 0.26 	&	$i$	&	LCO	 \\
4.668	&	20.64 $\pm$ 0.14 	&	$i$	&	NOT	 \\
4.699	&	20.62 $\pm$ 0.08 	&	$z$	&	NOT	 \\
5.52	&	20.79 $\pm$ 0.03 	&	$r$	&	GTC	 \\
5.56	&	21.16 $\pm$ 0.16 	&	$i$	&	NOT	 \\
5.569	&	$>$ 21.20        	&	$z$	&	NOT	 \\
6.457	&	$>$ 21.00        	&	$z$	&	NOT	 \\
15.258	&	$>$ 21.40        	&	$i$	&	ALT	 \\
15.26	&	$>$ 20.60        	&	$z$	&	ALT	 \\
15.322	&	$>$ 20.70        	&	$g$	&	ALT	 \\
15.345	&	$>$ 19.70        	&	$i$	&	ALT	 \\
15.35	&	$>$ 18.90        	&	$z$	&	ALT	 \\
17.44	&	$>$ 22.10        	&	$i$	&	LCO	 \\
17.92	&	21.05 $\pm$ 0.15 	&	$J$	&	P200	 \\
18.582	&	22.24 $\pm$ 0.11 	&	$i$	&	NOT	 \\
19.553	&	22.43 $\pm$ 0.14 	&	$z$	&	NOT	 \\
29.3	&	$>$ 22.50        	&	$R$	&	AZT	 \\
29.3	&	$>$ 21.60        	&	$I$	&	AZT	 \\
30.5	&	$>$ 22.40        	&	$i$	&	NOT	 \\
43.65	&	$>$ 21.90        	&	$r$	&	SOAR	 \\
43.67	&	$>$ 22.65        	&	$g$	&	VST	 \\
43.69	&	$>$ 22.30        	&	$r$	&	VST	 \\
43.72	&	23.14 $\pm$ 0.19 	&	$i$	&	VST	 \\
49.94	&	25.98 $\pm$ 0.21 	&	$g$	&	Keck	 \\
49.94	&	24.19 $\pm$ 0.12 	&	$R$	&	Keck	 \\
49.96	&	23.54 $\pm$ 0.16 	&	$I$	&	Keck	 \\
49.97	&	25.70 $\pm$ 0.29 	&	$V$	&	Keck	 \\
\end{longtable}

\clearpage

\begin{table*}
\centering
\caption{Afterglow C-band flux density observed on May 3 by ATCA. Errors represent the 1$\sigma$ uncertainties.}
\label{table:flux_density}
\small
\begin{tabular}{cccc}
\toprule
Frequency & Flux Density & Expected Thermal Noise  & Synthesis beam \\
(MHz) & $(\mu$Jy) & ($\mu$Jy/beam) & (arcsec $\times$ arcsec)  \\
\midrule
5500 & 1011 $\pm$ 20 & 19.9 &40.55 $\times$ 1.79 \\ 
9000 & 1419 $\pm$ 17 & 16.8 &24.19 $\times$ 1.13\\
\bottomrule
\end{tabular}

\end{table*}

\end{document}